\documentclass[pra, twocolumn, amsmath,amssymb, superscriptaddress, nofootinbib, floatfix ]{revtex4-1}

\usepackage{multirow}
\usepackage{graphicx}
\usepackage{amsfonts,tikz}  
\usetikzlibrary{matrix}
\usepackage{epstopdf}
\usepackage[version=3]{mhchem}
\usepackage{color}
\usepackage{hyperref}
\usepackage{mathtools}
\usepackage{blindtext}
\usepackage{enumitem}

\hypersetup{%
  pdfpagemode=None, 
  pdfstartpage=1,
  pdfstartview=FitH,
  pdfmenubar=true,
  pdftoolbar=true,
  colorlinks = true,
  linkcolor=blue,
  citecolor=blue,
  urlcolor = blue,
  bookmarksopen=false
}

\newcommand{\Op}[1]{\ensuremath{\boldsymbol{\mathsf{\hat{#1}}}}}
\newcommand{\op}[1]{\ensuremath{\mathsf{\hat{#1}}}}

\newcommand{\Bra}[1]{\ensuremath{\left\langle #1 \right\vert}}
\newcommand{\KetBra}[2]{\Ket{#1}\kern-0.1em\Bra{#2}}
\newcommand{\Ket}[1]{\ensuremath{\left\vert #1 \right\rangle}}

\definecolor{Pantone268}{cmyk}{0.82,1.0,0.0,0.12}
\definecolor{HunterOrange}{cmyk}{0.0,0.55,1.0,0.0}

\begin{document}

\title{Quantum control of photoelectron circular dichroism}
\date{\today}

\author{R.~Esteban Goetz}
\affiliation{Department of  Physics, Kansas State University, 116 Cardwell Hall, 
1228 N. 17th St. Manhattan, KS 66506-2601 }

\author{Christiane P. Koch}
\affiliation{Theoretische Physik, Universit\"{a}t Kassel, Heinrich-Plett-Str. 40, 
D-34132 Kassel, Germany}

\author{Loren Greenman}
\email{lgreenman@phys.ksu.edu}
\affiliation{Department of Physics, Kansas State University, 116 Cardwell Hall, 
1228 N. 17th St. Manhattan, KS 66506-2601 }

\begin{abstract}

We demonstrate coherent control over the photoelectron circular dichroism           
in randomly oriented chiral molecules, based on quantum interference between        
multiple photoionization pathways.  
To significantly enhance the chiral signature, we use a finite manifold of          
indistinguishable (1+1') REMPI pathways                                             
interfering at a common photoelectron energy but probing different                  
intermediate states.                                                                
We show that this coherent control mechanism maximizes the number of molecular  states that constructively contribute to the dichroism at an optimal            
photoelectron energy and thus 
outperforms other schemes,             
including interference between opposite-parity pathways driven by bichromatic       
($\omega,2\omega$) fields as well as sequential pump-probe ionization.

\end{abstract}

\maketitle

Chiral molecules are non-superimposable mirror images of each other, referred to
as enantiomers. Recent advances in measuring enantiomer-sensitive observables in
gas phase table-top
experiments~\cite{LuxAngewandte12,PattersonNature13,PitzerScience13,HerwigScience13}
have brought chiral molecules into the spotlight of current AMO research. One of
these observables is photoelectron circular dichroism (PECD), i.e., the
differential photoelectron angular distribution obtained by ionizing randomly
oriented molecules with left circularly and right circularly polarized
light~\cite{LuxAngewandte12,BoeweringPRL01,MeierhenrichACIE2010,LehmannJCP13,CombyJPCL16,BeaulieuSci17}.
PECD is a purely electric dipole effect, yielding much stronger signals than
traditional absorption circular dichroism (CD), which involves 
also the magnetic dipole of the probed transition. 
It can be quantified  
by the odd-moment coefficients
in the expansion of the  photoelectron angular distribution 
into Legendre polynomials.
The simplest explanation for PECD is provided by perturbation theory for one-photon ionization~\cite{RitchiePRA76}: It is the small difference in magnitude between dipole matrix elements with opposite sign $m$ quantum number, occurring only for chiral molecules, that results in a net effect when averaging over all molecular orientations. 
More intuitively, two non-parallel vectors are needed to provide an orientation
with which to probe the handedness of the molecular scaffold and create
a pseudo-scalar observable. While in
traditional CD these are the electric and magnetic dipole moment, the
photoelectron momentum provides the second vector  in PECD. This picture
connects PECD with the general framework for electric--dipole-based chiral observables~\cite{Ordonez18}. 
Perturbation theory can also explain the PECD observed in resonantly enhanced multi-photon ionization (REMPI)~\cite{LuxAngewandte12}, in terms of the electronically excited intermediate state of the REMPI process~\cite{GoetzJCP17}.
Dependence of the chiral signal on excitation wavelength is 
then understood in terms of probing different intermediate states~\cite{KastnerJCP17}. Whether PECD is amenable to 
coherent control by suitably shaping the 
ionizing pulses 
is an open question~\cite{Wollenhaupt2016NJP}. 

Here, we address this question by making use of optimal control theory
and show that, for a 
chiral methane derivative, \ce{CHBrClF}, quantum interference 
between distinct two-photon ionization pathways                            
significantly enhances PECD.
To this end, we combine a many-body description                            
of the electron dynamics,
scattering theory to efficiently describe the photoelectron
continuum~\cite{GianturcoJCP94,NatalenseJCP99,Greenman2017variational},
and second-order time-dependent perturbation theory
with an optimization technique~\cite{GoetzSpa2016}.
We use this approach                      
to maximize the PECD for \ce{CHBrClF} 
while fully accounting for the chiral nature of the potential experienced by the photoelectron. 
We use \ce{CHBrClF} as one of the simplest chiral molecules that has featured prominently in recent experiments~\cite{PitzerScience13} 
but expect our findings to be relevant for larger molecules as well.

We first detail our methodology to calculate the photoelectron spectrum and PECD. 
Keeping the nuclei fixed and neglecting relativistic effects,                           
the  Schr\"odinger equation for the many electron system reads 
\begin{eqnarray}
 \label{eq:SCHCIS0}
 i\dfrac{\partial}{\partial\,t}|\Psi^N(t)\rangle  &=& \Big[\op{H}_0 + \op{H}_1 - \boldsymbol{\mathcal{E}}(t)\cdot\Op{r}\Big]|\Psi^N(t)\rangle\,,
 \end{eqnarray}
 where $\op{H}_0$ and $\op{H}_1$ refer to the mean-field Fock operator and the
 residual Coulomb interaction, respectively. Accounting for one-particle one-hole excitations only, the many-body
  wave function is described by the manifold~\cite{KlamrothPRB03}
\begin{eqnarray}
\label{eq:cis_expansion} 
|\Psi^N(t)\rangle &=&
\alpha_0(t)\,e^{-i\varepsilon_{o}t}\,|\Phi_0\rangle +\sum_{i,a}\alpha^{a}_{i}(t)\,e^{-i\varepsilon_i^a t}\,|\Phi^{a}_{i}\rangle\\\nonumber
&&\quad\quad\quad\quad\quad+\sum_{i}\int d\boldsymbol{k}\,\alpha^{\boldsymbol{k}}_{i}(t)\,e^{-i\varepsilon_i^k t}\,|\Phi^{\boldsymbol{k}}_{i}\rangle\,,
\end{eqnarray}
where $\alpha_0(t)$, $\alpha^a_i(t)$ and $\alpha^{\boldsymbol{k}}_{i}(t)$ are time-dependent coefficients, and 
$|\Phi_0\rangle$ 
refers to the Hartree-Fock ground state.
$|\Phi^{a}_{i}\rangle = \Op{c}^\dagger_a\Op{c}_i|\Phi_0\rangle$ and 
$|\Phi^{\boldsymbol{k}}_{i}\rangle$ describe
one-particle one-hole excitations from an initially occupied
orbital $i$ to a bound unoccupied orbital $a$ or a continuum state 
with energy $\vert\boldsymbol{k}\vert^2/2$.  
\begin{figure}[!tb]
  \centering
  \includegraphics[width=0.99\linewidth]{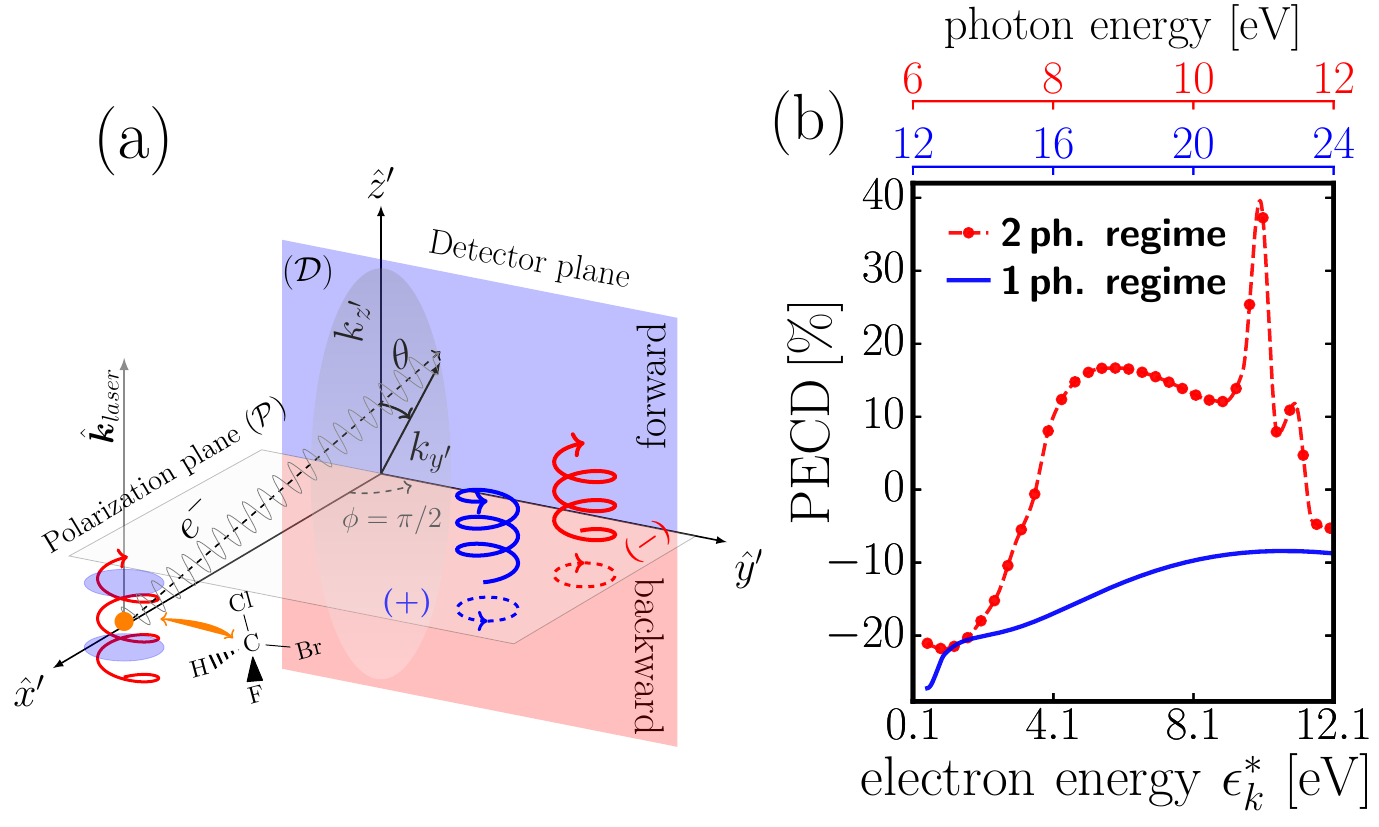}            
     \caption{                          
  (a) A randomly oriented ensemble of \ce{CHBrClF}
  molecules (orange ball) is
  ionized with left $(+)$ or right $(-)$  circularly polarized light, and the emitted electron  is measured in the $(z^\prime,y^\prime)$-plane.
  The polarization plane $(\mathcal{P})$ defines the $(x^\prime,y^\prime)$-plane and
  the vector $z^\prime$                              
normal to $\mathcal{P}$ is given by             
    the laser propagation direction. (b) Maximum PECD
    over all angles as a function 
   of the photoelectron energy  after ionization by a 
Gaussian pulse with central frequency $\omega$ denoted in terms of photon
  energy.
   The photon order is determined using the anisotropy parameters in Eq.~\eqref{eq:pecd.final}. 
  }
\label{fig:Figure1}
\end{figure}
To model an ensemble of randomly oriented 
molecules, we average                                                           
over all Euler angles $\gamma_{\mathcal{R}}=(\alpha,\beta,\gamma)$, see supplemental material~\cite{SupplementalMaterial} for detailed description, which also contains
Refs.~\cite{szabo2012modern,baertschy2001accurate,miller1987new,GianturcoJCP94,NatalenseJCP99, Greenman2017variational,RohringerPRA06,edmonds2016angular,GoetzSpa2016, werner2012molpro,werner2012molprowires}. 
The orientation-averaged photoelectron momentum distribution is 
obtained upon integration over $\gamma_{\mathcal R}$ and 
incoherent summation over the initially occupied contributing orbitals $i$ 
in the Hartree-Fock ground state, 
\begin{eqnarray}
 \label{eq:ExactForm}
 \dfrac{d^2\sigma}{d\epsilon_k\,d\Omega_{\boldsymbol{k}^\prime}} &=&  
 \sum_{i\in\text{occ}}\int |\alpha_i^{\boldsymbol{k}^\prime}(t;\gamma_{\mathcal{R}})|^2\, d^3\gamma_{\mathcal{R}}\,, 
\end{eqnarray} 
for $t\rightarrow\infty$ and
with $\boldsymbol{k}^\prime$ denoting the
momentum measured in the laboratory frame, defined by the propagation 
direction of the light beam along $z^\prime$, as indicated in Fig.~\ref{fig:Figure1}(a). 
The photoionization process is 
captured by the coefficients $\alpha^{\boldsymbol{k}^\prime}_i(t;\gamma_{\mathcal{R}})$.
It requires an accurate description of the scattering portion of the
wave function, which presents a formidable computational challenge
for a many-electron system with no symmetry. To 
reduce the computational cost, 
we                           
resort to solving Eq.~\eqref{eq:SCHCIS0} perturbatively. 
A second-order                                        
treatment allows us to manipulate quantum interferences
between  conventional
opposite-parity,  
as well as 
same-parity (two-photon)  pathways.
These interferences can be exploited to control the
differential and integral cross section in systems with no inversion center of
symmetry~\cite{BrumerReportsPhys,shapiro2012quantum}. 
Restricting the maximum field amplitude
and the ionization yield to ensure 
the validity of the perturbation approximation,                        
Eq.~\eqref{eq:ExactForm}      
 simplifies to 
\begin{eqnarray}
 \label{eq:SecondOrderApprox}
 \dfrac{d^2\sigma}{d\epsilon_k\,d\Omega_{\boldsymbol{k}^\prime}} &\approx&  \int \big|\alpha^{\boldsymbol{k}^\prime\,(1)}_{i_0}(t;\gamma_{\mathcal{R}}) +\alpha^{\boldsymbol{k}^\prime\,(2)}_{i_0}(t;\gamma_{\mathcal{R}})  \big|^2 d^3\gamma_{\mathcal{R}}\,,\quad
\end{eqnarray} 
for $t\to\infty$ and with $\alpha^{\boldsymbol{k}^\prime\,(1,2)}_{i_0}(t;\gamma_{\mathcal{R}})$, 
the first, resp. second, order correction~\cite{SupplementalMaterial}.
Second-order terms             
account for two-photon ionization pathways, from $i_0$ to $\boldsymbol{k}^\prime$ via different unoccupied  orbitals $a$. 

We restrict the electron dynamics to be influenced by the mean-field molecular electrostatic potential and time-dependent field  
only. 
The orbitals participating during the photoionization are described 
by the manifold of the HOMO (labeled $i_0$, with ionization potential $\omega_0$) and unoccupied orbitals  
defined by the eigenfunctions of the field-free Fock operator together with the scattering states
defined by the excitation $|\Phi^{\boldsymbol{k}^\prime}_{i_0}\rangle$. 
\begin{figure*}[!tb]
\centering
\includegraphics[width=0.90\linewidth]{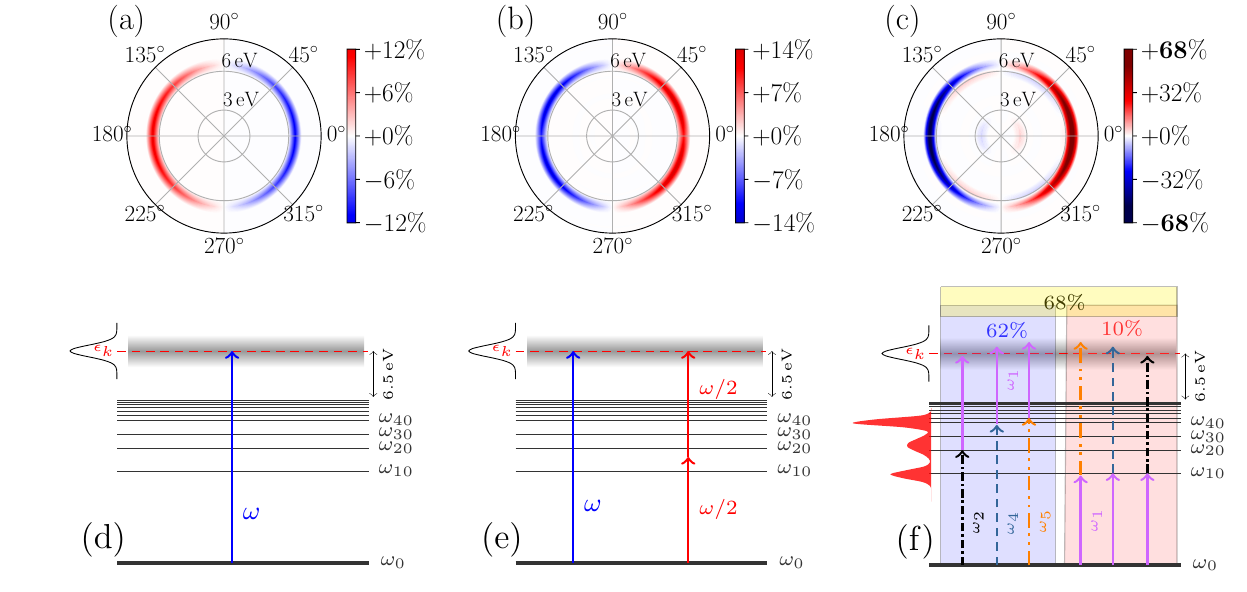} 
  \caption{(a-c): Angularly resolved PECD obtained with (a) a monochromatic reference field driving one-photon ionization, (b) an optimized bichromatic
  $(\omega,2\omega)$ pulse, and (c)  
making use of interference in even-parity two-photon pathways. 
(d-f): Corresponding ionization schemes.
The control mechanism for              
two-photon pathway interference (f)
is based on   
probing different intermediate states that interfere constructively at a common
continuum photoelectron energy within the spectral bandwidth. Restricting the 
control mechanism to the pump-probe scenario with time-delayed pulses results in a PECD of 62\%, resp. 10\%, whereas pulses overlapping in time realize the maximum PECD of 68\%. 
}
\label{fig:Figure2}
\end{figure*}
The Hartree-Fock orbitals were obtained using
the \texttt{MOLPRO}~\cite{werner2012molpro,werner2012molprowires} program
package with the \texttt{aug-cc-pVDZ} basis set~\cite{augccpvdz}.
The scattering portion $\varphi^{-}_{\boldsymbol{k}}(\boldsymbol{r})$ of the total wave function is an eigenfunction of the scattering problem, 
\begin{eqnarray}
  \label{eq:epolyscat}
  \left[-\dfrac{\nabla^2}{2} - \dfrac{1}{r} + \Op{V} - \dfrac{k^2}{2} \right]\varphi^{-}_{\boldsymbol{k}}(\boldsymbol{r})   &=&0 \,,
\end{eqnarray}
where $\Op{V}(\boldsymbol{r})$ is the short-range part of the electron-ion
interaction. Equation~\eqref{eq:epolyscat}  
is solved using
a locally modified version of the \texttt{ePolyScat} program 
package~\cite{GianturcoJCP94,NatalenseJCP99,Greenman2017variational}, see
the supplemental material~\cite{SupplementalMaterial} for more detail. PECD is calculated by expanding Eq.~\eqref{eq:SecondOrderApprox}
into Legendre polynomials
$P_\ell^m$, 
 \begin{eqnarray}
  \label{eq:CrossX}
  \dfrac{d^2\sigma^{(\pm)}}{d\epsilon_k\,d\Omega_{\boldsymbol{k}^\prime}} &=&
      \sum_{\ell,m} \beta^{(\pm)}_{\ell,m}(\epsilon_k)\,P^{m}_{\ell}(\cos\theta^\prime)\, e^{i\,m\varphi^\prime}\,,
 \end{eqnarray}
 where $\pm$ distinguishes the momentum distribution obtained with left 
 $(+)$  and right $(-)$ circularly polarized  light. 
 The anisotropy  parameters $\beta^{(\pm)}_{\ell,m}(\epsilon_k)$ are decomposed
 into contributions from the one- and two-photon ionization 
 pathways and their interference,  
 \begin{eqnarray}
 \label{eq:decomposition}
 \beta^{(\pm)}_{\ell,m}(\epsilon_k) &=& \beta^{(\pm)\, 1ph}_{\ell,m}(\epsilon_k)\, +\, \beta^{(\pm)\, 2ph}_{\ell,m}(\epsilon_k)\, +\, \beta^{(\pm)\,\text{int}}_{\ell,m}(\epsilon_k)\,.\quad 
 \end{eqnarray}
PECD is the non-vanishing component  
that remains after subtracting Eq.~\eqref{eq:CrossX} obtained with left and right 
circularly polarized light~\cite{LuxAngewandte12,BoeweringPRL01,MeierhenrichACIE2010,LehmannJCP13,CombyJPCL16,BeaulieuSci17} 
and reads~\cite{SupplementalMaterial}, for $\phi=\pi/2$, 
\begin{eqnarray}
\label{eq:pecd.final}
\text{PECD}(\epsilon_k,\theta,\phi=\pi/2) &=& 2\sum_{n,k}\,\beta^{(+)\,n ph }_{2k+1,0}(\epsilon_k)\,\text{P}^0_{2k+1}(\cos\theta)\nonumber\\
                                          && +6\, \mathrm{Im}\big[\beta^{(+)\text{int}}_{2,1}(\epsilon_k)\big]\,\sin(2\theta)\,.
\end{eqnarray}
Anisotropy parameters and PECD are expressed  
in percentage of the peak 
photoelectron intensity.  
The driving electric field  is 
parametrized,  
\begin{eqnarray}
  \label{eq:parametrization}
\epsilon(t)&=&\sum_{j=1}\epsilon_j\,e^{-(t-\tau_j)^2/2\sigma_j^2}\cos(\omega_j(t-\tau_j) + \phi_j)\,, 
\end{eqnarray}
with       
$\epsilon_j,\,\omega_j, \phi_j$  
the amplitude,
frequency, and carrier envelope phase of the $j$th pulse with full width at half 
maximum $\text{FWHM}=2\sqrt{2\ln{2}}\sigma_j$
and time delay $\tau_j$, 
which are optimized following Ref.~\cite{GoetzSpa2016}. To ensure the validity
of the perturbation approximation, we constrain the maximal
peak intensity to values not exceeding $1.0\times 10^{11}\,$W/cm$^2$,  
which was found to be an appropriate upper limit in bichromatic photoionization studies~\cite{douguet2016photoelectronCirc}.  

We first resolve the PECD as a function of the photon energy using a 
$25\,$fs (FWHM) monochromatic laser field (single frequency component in Eq.~\eqref{eq:parametrization}) with peak intensity $I_0=5\times 10^{10}\,$W/cm$^2$.
The resulting single- and two-photon PECD as a function of the photoelectron
energy are
shown in Fig.~\ref{fig:Figure1}(b). Figure~\ref{fig:Figure2}(a) shows the
angularly resolved PECD for a photon energy of $18.4\,$eV. We now address the question whether the PECD can be enhanced by allowing for more ionization pathways including their interference to contribute.
  
The use of quantum interference between one- and two-photoionization pathways  is a general, well-documented 
control mechanism~\cite{BrumerReportsPhys,shapiro2012quantum,douguet2017above,douguet2017above,douguet2016photoelectronCirc}, and bichromatic 
pulses have been suggested to realize this scenario for atoms
using  linearly~\cite{douguet2017above,douguet2017above}                               
and circularly~\cite{douguet2016photoelectronCirc}                          
polarized light. Control of anisotropy  after bichromatic ionization is also predicted for randomly oriented 
chiral molecules~\cite{Demekhin2018}. In this letter, we demonstrate, however, that interference between distinct two-photon ionization pathways 
results in a more efficient control mechanism to maximize PECD.

To this end, we  first optimize driving fields constraining the frequency
components to bichromatic $(\omega,2\omega)$ pulses. 
The PECD resulting from the optimized bichromatic $(\omega, 2\omega)$ pulse
reaches a maximum  
$\mathrm{PECD}$ of $20\%$  at a photoelectron energy of $10$ eV. This is comparable to asymmetries predicted 
for $(\omega,2\omega)$ bichromatic fields, linearly polarized in two mutually-orthogonal directions 
employing rotationally tailored laser pulses for control~\cite{Demekhin2018}. 
In a second step, we allow 
complete freedom for the photon energies of the driving field. 
With a  maximal peak intensity of 
$3.5\times 10^{10}\,$W/cm$^2$ (for a total ionization yield of 6\%), the  fully optimized field              
is found to significantly enhance the PECD to $68\%$.                           
The corresponding photoelectron spectrum peaks
at an energy of $6.5\,$eV.

\begin{figure*}[!tb]
\centering
\includegraphics[width=0.99\linewidth]{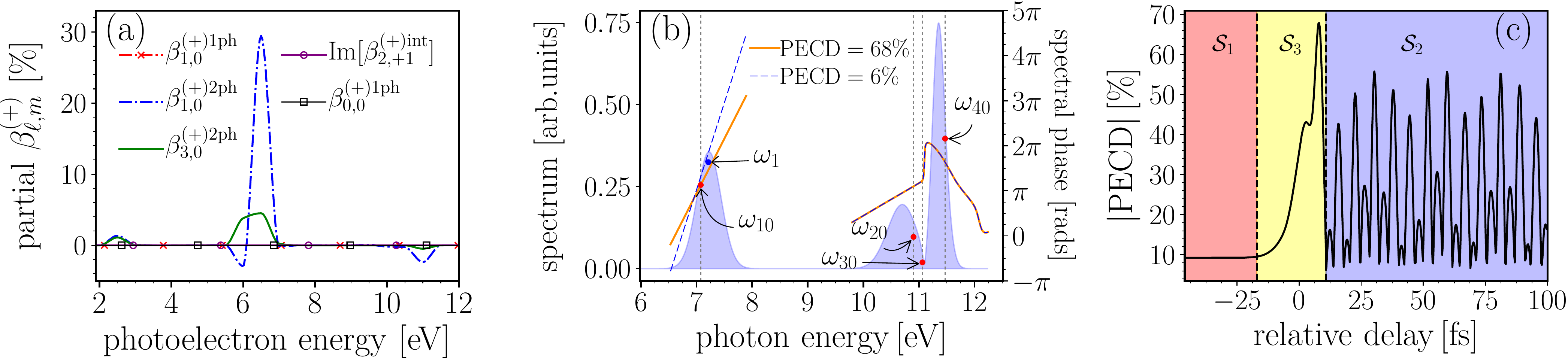}          
  \caption{Anisotropy parameters (a) 
for the fully optimized field as used in Fig.~\ref{fig:Figure2}(c).                                                             
Spectrum (blue)
and spectral phase                                      
  (solid orange line) of the fully optimized pulse are shown in (b). The frequencies $\omega_{j0}$ denote  
the transition energies
 between the $\mathrm{HOMO}$ and $\mathrm{LUMO}+j-1$ orbitals.
Modifying the spectral phase (orange solid vs blue dashed lines)  
while keeping the spectrum unchanged dramatically alters the PECD.
  PECD 
  as a function of the time delay (c) between the $\omega_1$-component 
  and the higher frequency components of the optimized pulse shown in panel (b). 
  In regions $\mathcal{S}_1$ and $\mathcal{S}_2$, the subpulses 
  are temporally separated (pump-probe scenario) with 
  $\mathcal{S}_1$ corresponding to ionization via the LUMO only 
  (cf. right part of Fig.~\ref{fig:Figure2}(f)) whereas
in    $\mathcal{S}_2$, ionization proceeds via a superposition of different excited states without the LUMO (cf. left part of Fig.~\ref{fig:Figure2}(f)). In $\mathcal{S}_3$, 
pump and probe pulses overlap in time such that interference between all two-photon ionization pathways can be exploited.
} 
\label{fig:Figure3}
\end{figure*}
PECD is known to strongly depend on the final continuum states.  It is
therefore important to  disentangle the kinetic energy effects, i.e., contributions arising from the final continuum 
state --here with energy $6.5\,$eV and $10\,$eV                              
-- 
and those from different photoionization pathways leading to the same final state.                   
We therefore compare the PECD obtained with a reference field driving one-photon
ionization,                         
the optimized bichromatic
$(\omega,2\omega)$ and fully optimized pulses                                              
resulting in the same photoelectron
kinetic energy.    
                 
The results for a photoelectron kinetic energy of $6.5\,$eV are shown in Fig.~\ref{fig:Figure2}, whereas those for $10\,$eV are found in the supplemental material~\cite{SupplementalMaterial}.
At a photoelectron energy of $6.5\,$eV, the PECD is enhanced from 12\% using
one-photon ionization to 14\% using an interfering two-photon ionization
pathway, i.e., bichromatic control, whereas
a single two-photon ionization pathway that don't include interference  
reaches 16.2\% PECD at $6.5\,$eV as shown in Fig.~\ref{fig:Figure1}(b). 
                 
All of these are significantly smaller than 68\% obtained for the fully optimized pulse where only two-photon pathways, but many more of them, cf. Fig.~\ref{fig:Figure2}(f), interfere. This picture holds also at a photoelectron kinetic energy of 10$\,$eV, where the 
maximum PECD for the reference field, optimized bichromatic $(\omega,2\omega)$ and fully-optimized pulses 
amounts to $8.5\%$, $20\%$ and $64\%$, respectively~\cite{SupplementalMaterial}.

The control mechanism for the fully optimized field of Fig.~\ref{fig:Figure2}(f) is further analyzed in  
Fig.~\ref{fig:Figure3} with Fig.~\ref{fig:Figure3}(a) showing the different anisotropy parameters 
for the fully optimized pulse. In contrast to the reference and optimized bichromatic $(\omega,2\omega)$
pulses, neither single-photon ionization nor interference between one- and two-photon pathways
contribute for the fully optimized pulse.
In fact, both $\beta^{(+)\mathrm{1ph}}_{1,0}$ and $\beta^{(+)\mathrm{int}}_{2,+1}$ vanish 
over the entire energy domain. Furthermore, the one-photon ionization pathway is  
completely suppressed, even for the symmetric part,   
since  $\beta^{(+)\mathrm{1ph}}_{0,0}= 0$. Instead, the remarkable 
enhancement of the PECD is indeed solely due 
to even-parity,  
i.e., two-photon ionization pathways  because
both $\beta^{(+)\mathrm{2ph}}_{1,0}$ and $\beta^{(+)\mathrm{2ph}}_{3,0}$ are non-zero.  

Analyzing the spectrum of the fully optimized field, 
shown in 
Fig.~\ref{fig:Figure3}(b), we identify the interference between different two-photon ionization pathways to give rise to the control observed in Fig.~\ref{fig:Figure2}(c).
The spectrum contains peaks  
at $\omega_1 = 7.24\,$eV, 
$\omega_2 =10.70\,$eV, 
and $\omega_3= 11.35\,$eV  
and overlaps with  
transitions from the 
HOMO to the first unoccupied orbitals, 
namely, $\omega_{10} = 7.07\,$eV, $\omega_{20}=10.90\,$eV, $\omega_{30}=11.06\,$eV and
$\omega_{40}=11.47\,$eV, which need to be compared to the calculated ionization threshold of  11.88$\,$eV.
The two-photon ionization pathway  $2\times\omega_{10}$
promotes the photoelectron to $2.27\,$eV 
and explains  
the small peak at $2.39\,$eV in Fig.~\ref{fig:Figure3}(a). Conversely,
the pathways
$2\times\omega_2$, $\omega_2+\omega_3$, $\omega_2+\omega_4$, $\dots$,
explain the small PECD at $11\,$eV.  However, these pathways do not contribute 
much to PECD which is mainly due to the peak at $6.5\,$eV.                           
The most important pathways leading to 6.5$\,$eV, cf. Fig.~\ref{fig:Figure1}(f),  are $\omega_{20}+(\omega_{10}+\delta\omega_1)$ and similarly $(\omega_{10}+\delta\omega_1) +\omega_{20}$ with an offset of $\delta\omega_1 = 0.4\,$eV as well as 
$\omega_{40} + (\omega_{10}-\delta\omega)$ with $\delta\omega=0.12\,$eV. The latter probes the LUMO+3, 
whereas the former two probe the LUMO+1 and LUMO. The required offsets are available within the spectral bandwidth. 
The pathways $\omega_{30} + \omega_{10}$ and $\omega_{10} + \omega_{30}$,          
probing the LUMO+2, are also compatible with the pulse spectrum; however, 
the frequency $\omega_{30}$ is suppressed, and removing the $\mathrm{LUMO+2}$  
decreases the  PECD by only $0.4\%$.
In other words, the high frequency components of the optimized field 
correspond to  
photon energies  
which resonantly excite the first $\text{LUMO}+j$ 
orbitals $(j=1,2,\dots)$, while the 
peak centered at $\omega_1$ can either excite the LUMO from the ground state or ionize the LUMO$+j$ population, cf. Fig.~\ref{fig:Figure2}(f).
Its bandwidth guarantees 
interference at a common photoelectron energy.
Thus, the 
width and peak position at $\omega_1$ are key for the 
constructive interference among a finite manifold of two-photon ionization 
pathways at a common final photoelectron energy to significantly enhance  PECD.  
Constraining $\omega_1$ to be exactly $\omega_{10}$ while reducing its spectral bandwidth results 
in a smaller PECD  
$(\approx 50\%)$. Conversely, allowing $\omega_1$ to be further
blue-shifted  
with respect to $\omega_{10}$ while increasing the spectral bandwidth such that it still overlaps with $\omega_{10}$ results
in a  $\mathrm{PECD}$ of about 70\% (not shown due to the large bandwidth of the field).  

The coherent nature of the control mechanism is further confirmed  
by modifying the spectral phase of the optimized pulse 
while keeping the spectral amplitude  
unaffected, cf. dashed lines in Fig.~\ref{fig:Figure3}(b).
This corresponds to introducing a time delay between the high and low-frequency components of the pulse. 
Figure~\ref{fig:Figure3}(c) shows the variation of the PECD, between
$68\%$ to $6\%$, as a function of this 
time delay.                                                                                                      
Positive (negative) delays correspond to the high-frequency components arriving before (after) the low-frequency components, as 
verified by inspecting the Wigner distribution function of the pulses. 
For negative time delays,                             
highlighted in red in Figs.~\ref{fig:Figure2}(f) and~\ref{fig:Figure3}(c), only the LUMO is excited. 
PECD thus does not depend on the time delay and reflects the chiral signature of the LUMO only, which amounts to about 10\%.
For positive time delays, highlighted in blue in Figs.~\ref{fig:Figure2}(f) and~\ref{fig:Figure3}(c), the high-frequency components of the pulse 
prepare a superposition of higher excited states, such that the PECD depends on the time delay and contains the chiral fingerprints of the LUMO$+j$ ($j\geq 1$) with a maximum  of 55\%. 
These two scenarios correspond to 
pump-probe control~\cite{TannorJCP85,TannorJCP86} where 
the pump pulse spectrum selects the manifold of intermediate states that 
contribute. PECD can be pushed to 62\% by further optimization of time-separated pump and probe pulses for positive delays.
However, the maximal value of PECD, 68\%, is obtained when pump and probe overlap, as highlighted in yellow in Figs.~\ref{fig:Figure2}(f) and~\ref{fig:Figure2}(c). 
This can be rationalized by exploiting interference of all the pathways, including 
the two-photon ionization through the LUMO, depicted  in Fig.~\ref{fig:Figure2}(f).

In conclusion, we have identified constructive interference in
two-photon photoionization to significantly enhance PECD of randomly
oriented CHBrClF molecules. Control is achieved 
via various (1+1$^\prime$) REMPI pathways
leading to a common final photoelectron state but probing different intermediate states. Separating pump and probe photons in time slightly reduces the number of pathways that may interfere and thus the PECD. 
In this  excitation scheme based on interference of same-parity pathways, we find significantly larger PECD than can be obtained                     
with optimized  bichromatic circularly polarized fields where opposite parity pathways are made to interfere constructively.
It will be straightforward to extend this type of control to molecules other than CHBrClF, with only the central frequencies and spectral widths depending on the specific chiral molecule. 
Higher-order terms in the perturbation expansion, while requiring
larger amplitudes, are likely to facilitate even more pathway
interference and could also be used to drive photoionization with
optical instead of XUV pulses. Whether an upper bound to PECD exists and what type of driving field would  saturate it is yet unknown.

We would like to thank  Thomas Baumert for helpful comments on the manuscript.
The computing for this project was performed on the Beocat Research Cluster at Kansas State University, which is funded in part by NSF grants CNS-1006860, EPS-1006860, and EPS-0919443.
CPK acknowledges financial support from the Deutsche Forschungsgemeinschaft (CRC
    1319).

\onecolumngrid
\cleardoublepage
\twocolumngrid
\setcounter{page}{1}
\thispagestyle{empty}

\onecolumngrid
  \subsection*{\large \textbf{\sffamily{Supplemental material for\\[0.2cm] Quantum control of photoelectron circular dichroism}}}
\twocolumngrid
\onecolumngrid
\begin{center}
\small
  R.~Esteban Goetz,$^1$ Christiane~P. Koch,$^2$ and Loren Greenman$^{1,\textcolor{blue}{*}}$\\[0.2cm]
  \textit{$^1$Department of  Physics, Kansas State University,\\
116 Cardwell Hall, 1228 N. 17th St. Manhattan, KS 66506-2601}\\ 
\textit{$^2$Theoretische Physik, Universit\"{a}t Kassel, Heinrich-Plett-Str. 40, D-34132 Kassel, Germany}\\
(Dated: December 16, 2018)\\[0.8cm]
\end{center}
\twocolumngrid

\setcounter{equation}{0}
For completeness, we provide here the details of the derivation of the
orientation-averaged experimental observables, i.e., the anisotropy
parameters and photoelectron circular dichroism (PECD), within the framework of 
second-order time-dependent
perturbation approximation.  
Section~\ref{sec:section1} introduces the equations of motion for the electron
dynamics and describes the perturbation treatment
of the light-matter interaction. It also summarizes
the operations  needed to express the photoelectron momentum
distribution in the laboratory
frame of reference. Section~\ref{sec:section2} defines the
orientation averaged laboratory-frame
anisotropy parameters for  the one-photon and two-photon ionization pathways and their
interference. In
Sec.~\ref{sec:section3} the symmetry properties of the different anisotropy
parameters under polarization reversal are analyzed, and the PECD is defined
accordingly. Section~\ref{sec:ControlPecd} presents the optimization algorithm and
cost functional.  

Section~\ref{sec:HemisphereIntegrated} introduces 
the concept of ``hemisphere-averaged 
PECD'',  corresponding to a particular case of angle-integrated PECD,  
complementing the single- and two-photon PECD presented in Fig.~1(b) in the manuscript. 

Section~\ref{sec:DisantanglingOptimalInterference} provides the time-frequency analysis (Wigner
distribution function) of the fully-optimized pulse presented in the main text, 
allowing for disentangling the optimal multiple-two-photon ionization scheme
from all possible two-photon pathways, 
further extending the analysis presented in the manuscript. The time-frequency
analysis is accompanied by a more extended and detailed
discussion of the three possible scenarios defined by the time-delay
between the low and high frequency components of the fully optimized field
presented in the manuscript. 

Section~\ref{sec:ResultsAt10eV} complements Fig.~2 of the main text (with a final photoelectron energy of 6.5$\,$)  by showing the corresponding results at a different photoelectron energy  10$\,$eV, for which the largest PECD with bichromatic control is obtained.
Section~\ref{sec:ResultsAt10eV} also discusses the role of
the number of different two-photon ionization pathways probing different
orbitals but interfering at
a common photoelectron kinetic energy to maximize the PECD,  and provides
a                
study of the PECD as a function of the relative phase between
the fundamental and second harmonic for a bichromatic $(\omega,2\omega)$
pulse. Finally, sections~\ref{sec:section4} and~\ref{sec:section5} 
contain the input file parameters for the electronic structure and scattering
calculations, respectively.

\section{\textbf{\sffamily{Perturbation expansion}}}
\label{sec:section1}

\subsection{\textbf{\sffamily{Equations of motion}}}

Equations~(1) and (2) of the main text introduce the time-dependent many-electron Schr\"odinger equation and the expansion of the many-electron state $|\Psi^N(t)\rangle$ into the Hartree-Fock (HF) ground state and one-particle-one-hole excitations,  $|\Phi^a_i\rangle$.
The Slater determinant describing the HF ground-state is constructed from an anti-symmetric
product of one-electron spin-orbitals $\varphi_i$, 
\begin{eqnarray}
\label{eq:HForbs}
\Phi_0 = \mathcal{A}\displaystyle\prod^N_{i=1}\varphi_i(\mathbf{r}_i)
\end{eqnarray}
such that $\Op{H}_0|\varphi_p\rangle = \epsilon_p|\varphi_p\rangle$ where $\epsilon_p$ are the orbital energies. 
Expanding the many-electron Hamiltonian into the basis of HF ground state and one-particle-one-hole excitations, defining  
\begin{eqnarray}
  \varepsilon_0 &=&\langle\Phi_0|\Op{H}_0|\Phi_0\rangle\,,\nonumber\\                                     
    \varepsilon^a_i &=&\langle\Phi^a_i|\Op{H}_0|\Phi^a_i\rangle \nonumber\\
   &=& \epsilon_0
  + \epsilon_a-\epsilon_i\,, 
\end{eqnarray} 
and applying Brillouin's theorem~\cite{brillouin1933methode}, 
\begin{eqnarray}
  \quad\langle\Phi_0|\Op{H}_1|\Phi^a_i\rangle &=&0\,,
\end{eqnarray}
we arrive at 
a set of coupled equations for the expansion coefficients,
\begin{subequations}
\label{eq:coupled.set}
\begin{eqnarray}
  \label{eq:a0.diff}
  \dot{\alpha_0}(t) &=& i\boldsymbol{\mathcal{E}}(t)\Big[\langle\Phi_0|\Op{r}|\Phi_0\rangle\alpha_0(t)
\\
                    &&\quad\quad\quad+ 
\sum_{i,a}\langle\Phi_0|\Op{r}|\Phi^a_i\rangle e^{-i(\epsilon^a_i-\epsilon_0)t}\,\alpha^a_i(t)\Big]\nonumber\\
  \label{eq:aia.diff}
  \quad\,\dot{\alpha}^a_i(t) &=&  i\boldsymbol{\mathcal{E}}(t)\Big[
    \langle\Phi^a_i|\Op{r}|\Phi_0\rangle\,e^{-i(\epsilon_0-\epsilon^a_i)t} \alpha_0(t)\\
  &&\quad\quad\quad+ 
\sum_{j,b}\langle\Phi^a_i|\Op{r}|\Phi^b_j\rangle e^{-i(\epsilon^b_j-\epsilon^a_i)t}\,\alpha^b_j(t)\Big]\nonumber\\ &&-i                 
\sum_{j,b}\langle\Phi^a_i|\Op{H}_1|\Phi^b_j\rangle\alpha^b_j(t)e^{-(\epsilon^b_j - \epsilon^a_i)t}\,.\nonumber
\end{eqnarray}
\end{subequations}
In particular, for $a=\boldsymbol{k}$, the  
coefficients $\alpha^{\boldsymbol{k}}_{i}(t)$
describe the transition amplitude from an initially occupied 
orbital $i$ to a continuum state with energy $\epsilon_k = |\boldsymbol{k}|^2/2$ in the direction
$\boldsymbol{k}/|\boldsymbol{k}|$ with respect to the molecular frame of reference,
$\mathcal{R}$.
Similarly,  
$\alpha^{\boldsymbol{k}^\prime}_{i}(t)$ describe this transition
in the laboratory frame, $\mathcal{R}^\prime$.

\subsection{\textbf{\sffamily{Orientation-averaged momentum distribution}}}
\label{sec:orientation_averaged}

The  
orientation-averaged momentum distribution of photoelectrons with energy $\epsilon_k$
emitted within a solid angle $d\Omega_{\boldsymbol{k}^\prime}$ measured in
the laboratory frame  
is given by Eq.~(3) of the main text.
In order to  account for first- and second-order processes and their interference,
we solve the equation of motion for  the amplitudes
$\alpha^{\boldsymbol{k}^\prime}_i(t;\gamma_{\mathcal{R}})$ using second-order
time-dependent perturbation theory                                               
which results in Eq.~(4) of the main text.
We limit the calculations to the single-channel approximation
with  $i_0$ labeling the HOMO orbital.
The angle and energy resolved photoelectron distribution 
can then be written
in terms of the contributions from one- and two-photon ionization processes and
their interference,
\begin{eqnarray}
 \label{eq:SecondOrderApprox_split}
 \dfrac{d^2\sigma^{(\mu_0)}}{d\epsilon_k\,d\Omega_{\boldsymbol{k}^\prime}} &=&                                                                                                                                                                                        
 \dfrac{d^2\sigma^{(\mu_0)1ph}}{d\epsilon_k\,d\Omega_{\boldsymbol{k}^\prime}}+
 \dfrac{d^2\sigma^{(\mu_0)2ph}}{d\epsilon_k\,d\Omega_{\boldsymbol{k}^\prime}}+
 \dfrac{d^2\sigma^{(\mu_0)int}}{d\epsilon_k\,d\Omega_{\boldsymbol{k}^\prime}}\,,
\end{eqnarray} 
where $\mu_0 = (\pm 1,0)$ defines the spherical unit vector components of 
the polarization direction $\epsilon^\prime_{\mu_0}$ in $\mathcal{R}^\prime$.
It has been introduced to distinguish the photoelectron            
distribution 
obtained with left $(\mu_0=+1)$ and right $(\mu_0=-1)$ circularly polarized light or with 
linear $(\mu_0=0)$ polarization. The contribution from one-photon processes in Eq.~\eqref{eq:SecondOrderApprox_split} becomes 
\begin{eqnarray}
  \label{eq:1phcontr}
  \dfrac{d^2\sigma^{(\mu_0)1ph}}{d\epsilon_k\,d\Omega_{\boldsymbol{k}^\prime}}&=& \int\alpha^{(1)\boldsymbol{k}^\prime}_{i_0}(t;\gamma_{\mathcal{R}})\alpha^{*(1)\boldsymbol{k}^\prime}_{i_0}(t;\gamma_{\mathcal{R}})\,d^3\gamma_{\mathcal{R}}\\         
&=&
\sum_{L,M}\beta^{(\mu_0)1ph}_{L,M}(\epsilon_k)\,P^M_L(\cos\theta_{\boldsymbol{k}^\prime})\,e^{iM\phi_{\boldsymbol{k}^\prime}}\,,\quad \nonumber
\end{eqnarray}
where in the second line we have invoked an expansion into associate Legendre polynomials, $P^{M}_{L}(\cdot)$.
Similarly, contributions
to the photoelectron momentum distribution
originating from second-order order processes read
\begin{eqnarray}
  \label{eq:2phcontr}
  \dfrac{d^2\sigma^{(\mu_0)2ph}}{d\epsilon_k\,d\Omega_{\boldsymbol{k}^\prime}}&=& \int\alpha^{(2)\boldsymbol{k}^\prime}_{i_0}(t;\gamma_{\mathcal{R}})\alpha^{*(2)\boldsymbol{k}^\prime}_{i_0}(t;\gamma_{\mathcal{R}})\,{d^3\gamma_{\mathcal{R}}}\\         
&=&
\sum_{L,M}\beta^{(\mu_0)2ph}_{L,M}(\epsilon_k)\,P^M_L(\cos\theta_{\boldsymbol{k}^\prime})\,e^{iM\phi_{\boldsymbol{k}^\prime}}\,. \nonumber
\end{eqnarray}
For the interference terms between one- and two-photon ionization pathways
in Eq.~(4) of the main text, 
we define 
\begin{eqnarray}
  \label{eq:beta.int.def}
  \beta^{(\mu_0)int}_{L,M} &=&          
\int \alpha^{(1)\boldsymbol{k}^\prime}_{i_0}(\gamma_{\mathcal{R}})\alpha^{*(2)\boldsymbol{k}^\prime}_{i_0}(\gamma_{\mathcal{R}})\,d^3\gamma_{\mathcal{R}}\,. 
\end{eqnarray}
Using Eq.~\eqref{eq:beta.int.def}, the contribution from the interfering pathways to the photoelectron spectrum reads   
\begin{eqnarray}
  \label{eq:12phcontr}
  \dfrac{d^2\sigma^{(\mu_0)int}}{d\epsilon_k\,d\Omega_{\boldsymbol{k}^\prime}}&=& \int\Big( \alpha^{(1)\boldsymbol{k}^\prime}_{i_0}(t;\gamma_{\mathcal{R}})\alpha^{*(2)\boldsymbol{k}^\prime}_{i_0}(t;\gamma_{\mathcal{R}})  + c.c.\Big)\,d^3\gamma_{\mathcal{R}}\nonumber\\                
&=&\sum_{L,M}\Big(\beta^{(\mu_0)int}_{L,M}(\epsilon_k)\,e^{iM\phi_{\boldsymbol{k}^\prime}} + c.c.\Big)\, P^M_L(\cos\theta_{\boldsymbol{k}^\prime}),\nonumber
\end{eqnarray}
or, equivalently,
\begin{eqnarray}
  \label{eq:12phcontr.final}
  \dfrac{d^2\sigma^{(\mu_0)int}}{d\epsilon_k\,d\Omega_{\boldsymbol{k}^\prime}}&=& 2\sum_{L,M}\Big[\text{Re}\big[\beta^{(\mu_0)int}_{L,M}(\epsilon_k)\big]\cos(M\phi_{\boldsymbol{k}^\prime})\\          
&& -\, \text{Im}\big[\beta^{(\mu_0)int}_{L,M}(\epsilon_k)\big]\sin(M\phi_{\boldsymbol{k}^\prime})\Big]\,P^M_L(\cos\theta_{\boldsymbol{k}^\prime})\nonumber\,,
\end{eqnarray}
where the possible values of $L$ and $M$ are determined by $\mu_0$ as well as the symmetry properties after the orientation averaging. The portion 
of the photoelectron spectrum due to the interference term is  
sensitive to the relative phase between one- and two-photon processes. In particular,
measurements on the 
$(z^\prime,y^\prime)$-plane ($\phi_{\boldsymbol{k}^\prime}=\pi/2 $) and $(z^\prime,x^\prime)$-plane ($\phi_{\boldsymbol{k}^\prime}=0 $)  may allow  
reconstruction of the phase.
\subsection{\textbf{\sffamily{First-order corrections}}}

The first order correction of $\alpha_0(t)$ in Eq.~\eqref{eq:a0.diff} reads 
\begin{eqnarray}
  \label{eq:alpha_0.firstorder}
  \alpha^{(1)}_0(t) &=&
  i\langle\Phi_0|\Op{r}|\Phi_0\rangle\int_{-\infty}^{t}\boldsymbol{\mathcal{E}}(t^\prime)\,\alpha^{(0)}_0(t^\prime)\,dt^\prime\\
  && + i\sum_{i,a}\langle\Phi_0|\Op{r}|\Phi^a_i\rangle\int^t_{-\infty}
  e^{-i(\epsilon^a_i-\epsilon_0)t^\prime}\alpha^{a(0)}_i(t^\prime)\boldsymbol{\mathcal{E}}(t^\prime)\,dt^\prime\,,\nonumber
\end{eqnarray}
where $j$ in $\alpha_0^{(j)}(t)$ and $\alpha^{a(j)}_i(t)$ indicates the 
perturbation order. For convenience, both one-particle one-hole excitation to virtual bound and scattering 
states are contained in the index $a$ in Eq.~\eqref{eq:alpha_0.firstorder}. Using the zeroth order approximations $\alpha^{(0)}_0(t)\approx 1 $ and $\alpha^{a(0)}_i(t)\approx 0$, Eq.~\eqref{eq:alpha_0.firstorder} becomes   
\begin{eqnarray}
  \label{eq:alpha_0.firstorder00}
  \alpha^{(1)}_0(t) &=& i\langle\Phi_0|\Op{r}|\Phi_0\rangle\int_{-\infty}^{t}\boldsymbol{\mathcal{E}}(t^\prime)\,dt^\prime\,. 
\end{eqnarray}
The first order correction for the one-particle one-hole excitations reads 
\begin{eqnarray}
  \label{eq:alpha_ia.firstorder}
  \alpha^{a(1)}_i(t) &=& i\,\langle\Phi^a_i|\Op{r}|\Phi_0\rangle\int_{-\infty}^{t}\alpha^{(0)}_0(t^\prime)\,\,e^{-i(\epsilon_0-\epsilon^a_i)t^\prime}\,\boldsymbol{\mathcal{E}}(t^\prime)\,  dt^\prime\,\\ 
  && +\, i\sum_{j,b}\langle\Phi^a_i|\Op{r}|\Phi^b_i\rangle\int^t_{-\infty}\alpha^{b(0)}_i(t^\prime)\,e^{-i(\epsilon^b_j-\epsilon^a_i)t^\prime}\boldsymbol{\mathcal{E}}(t^\prime)\,dt^\prime\quad\nonumber\\         
  && -\,i \sum_{j,b}\langle\Phi^a_i|\Op{H}_1|\Phi^b_j\rangle\,\int^t_{-\infty} e^{-i(\epsilon^b_j-\epsilon^a_i)t^\prime}\alpha^{b(0)}_j(t^\prime)\,dt^\prime\,.\quad\nonumber
\end{eqnarray}
Using again $\alpha^{(0)}_0(t)\approx 1 $ and $\alpha^{a(0)}_j(t)\approx 0$,
Eq.~\eqref{eq:alpha_ia.firstorder} reduces to 
\begin{eqnarray}
  \label{eq:alpha_ia.firstorder2}
  \alpha^{a(1)}_i(t) &=&
  i\,\langle\Phi^a_i|\Op{r}|\Phi_0\rangle\int_{-\infty}^{t}e^{-i(\epsilon_0-\epsilon^a_i)}\boldsymbol{\mathcal{E}}(t^\prime)\,dt^\prime\,, 
\end{eqnarray}
where the two determinants differ only by                      
one spin-orbital. Application 
of the Slater-Condon rules for one-electron operators
$\mathcal O$~\cite{szabo2012modern}, 
\begin{eqnarray}
    \label{eq:SlaterCondom}
    \langle\Phi_0|\op{\mathcal{O}}|\Phi_0\rangle&=& \displaystyle\sum_i\langle\varphi_i|\op{\mathcal{O}}|\varphi_i\rangle\,\nonumber\\[0.0cm]
    \langle\Phi_0|\op{\mathcal{O}}|\Phi^a_i\rangle&=&\langle\varphi_i|\op{\mathcal{O}}|\varphi_a\rangle\,,                    
\end{eqnarray} 
to Eq.~\eqref{eq:alpha_ia.firstorder2} allows us to write the matrix elements in
terms of the HF one-electron spin-orbitals,
\begin{eqnarray}
  \label{eq:alpha_ia.firstorder2.slatter}
  \alpha^{a(1)}_i(t) &=& i\,\langle\varphi_a|\Op{r}|\varphi_i\rangle\int_{-\infty}^{t}e^{-i(\epsilon_0-\epsilon^a_i)}\boldsymbol{\mathcal{E}}(t^\prime)\,dt^\prime\,. 
\end{eqnarray}
with $\varphi_s$ the $s$th occupied or unoccupied  spin-orbital, cf.~Eq.~\eqref{eq:HForbs}.
In particular, the first-order correction for the quantity of interest, namely $\alpha^{\mathbf{k}}_{i_0}(t)$,  reads 
\begin{eqnarray}
  \label{eq:alpha_ik_first.preliminary}
  \alpha^{\boldsymbol{k}(1)}_{i_0}(t) &=&
  i\,\langle\Phi^{\boldsymbol{k}}_{i_0}|\Op{r}|\Phi_0\rangle\int_{-\infty}^{t}e^{-i(\epsilon_0-\epsilon^k_{i_0})}\boldsymbol{\mathcal{E}}(t^\prime)\,dt^\prime\,, 
\end{eqnarray}

Assuming no relaxation of the contributing orbitals, the   
total wave function $\Phi^{\boldsymbol{k}}_{i}(\boldsymbol{r}_N)$ can be
defined,
for any $i\in\text{occ}$, as an antisymmetrized
product, 
\begin{eqnarray}
  \Phi^{\boldsymbol{k}}_{i}(\boldsymbol{r}_1,\dots \boldsymbol{r}_N ) &=& 
\mathcal{A}_{N}\big[\varphi^-_{\boldsymbol{k}}(\boldsymbol{r}_N);\Phi_{i}(\boldsymbol{r}_{1},\dots  \boldsymbol{r}_{N-1})\big]\,,\quad
\end{eqnarray}
where $\varphi^-_{\boldsymbol{k}}(\boldsymbol{r}_N) $ is the scattering portion of
the wave function and  $\Phi_{i}( \boldsymbol{r}_{1},\dots\boldsymbol{r}_{N-1})$ the $N-1$ electron final
state after ionization. 
We obtain $\varphi^{-}_{\boldsymbol{k}}(\mathbf{r})$  
by solving the scattering problem 
\begin{eqnarray}
  \label{eq:epolyscat.supp}
  \left[-\dfrac{\nabla^2}{2} - \dfrac{1}{r} + \Op{V} - \dfrac{k^2}{2} \right]\varphi^{-}_{\boldsymbol{k}}(\mathbf{r})   &=&0 \,,
\end{eqnarray}
with scattering boundary conditions~\cite{baertschy2001accurate,miller1987new} for the outgoing wave $\varphi^{-}_{\boldsymbol{k}}(\mathbf{r})$ at
large distances $\boldsymbol{r}\rightarrow \infty$,  
and where $\Op{V}(\mathbf{r})$ describes the short-range part of the electron-ion
interaction. Equation~\eqref{eq:epolyscat.supp} including its matrix  and tensor elements  are computed using
a locally modified version of the \texttt{ePolyScat} program 
package~\cite{GianturcoJCP94,NatalenseJCP99,Greenman2017variational}. The orthogonality 
between the Hartree-Fock orbitals and scattering states
obtained from the scattering calculations have been numerically verified
within the tolerance range.

\subsection{\textbf{\sffamily{Second-order corrections}}}

The second order correction                                                                                                                
of $\alpha^{\boldsymbol{k}}_i(t)$ for $i\in\text{occ}$ is, according to Eq.~\eqref{eq:aia.diff}, 
\begin{eqnarray}
  \label{eq:a_ik.second}
  \alpha^{\boldsymbol{k}(2)}_i(t) &=&
  \,i\,\langle\Phi^{\boldsymbol{k}}_i|\Op{r}|\Phi_0\rangle\int^t_{-\infty}e^{i(\epsilon^k_i-\epsilon_0)t^\prime}\boldsymbol{\mathcal{E}}(t^\prime)\,\alpha^{(1)}_0(t^\prime)dt^\prime\\
  &&+  i\sum_{j,b}\langle\Phi^{\boldsymbol{k}}_{i_0}|\Op{r}|\Phi^b_j\rangle\int^t_{-\infty}e^{i(\epsilon^k_i-\epsilon^b_j)t^\prime}\boldsymbol{\mathcal{E}}(t^\prime)\,\alpha^{b(1)}_j(t^\prime)\,dt^\prime\nonumber\\
  &&- i\sum_{j,b}\langle\Phi^{\boldsymbol{k}}_i|\Op{H}_1|\Phi^b_j\rangle\int^t_{-\infty}e^{i(\epsilon^k_i-\epsilon^b_j)t^\prime}\,\alpha^{b(1)}_j(t^\prime)\,dt^\prime\nonumber\,.\\\nonumber
\end{eqnarray}
The electron-correlation effects  described by $\Op{H}_1$ will be
neglected. However, the exchange and Coulomb operator in $\Op{H}_0$  ensure the
excited electron to still experience a (chiral) Coulomb attraction to the residual
cation~\cite{RohringerPRA06}. To verify the chiral origin of PECD 
when $\Op{H}_1$ is ignored, we have
performed the same calculations using the achiral $\ce{CH4}$ and $\ce{N2}$ molecules, for which no PECD was
observed (below machine precision). 

Matrix elements are obtained using the rules for one-electron operators                
~\eqref{eq:SlaterCondom} together
with                                                 
\begin{eqnarray}
  \langle\Phi^a_i|\Op{r}|\Phi^b_j\rangle &=&\,\,\,\,\,\delta_{a,b}(1-\delta_{i,j})\,\langle\varphi_a|\Op{r}|\varphi_b\rangle\nonumber\\[0.1cm]
                                         &&+\,\delta_{i,j}(1-\delta_{a,b})\,\langle\varphi_j|\Op{r}|\varphi_i\rangle\nonumber\\[0.05cm] 
                                         &&+\,\delta_{i,j}\delta_{a,b}\,\sum_{\scriptsize r\in\{m^a_i\}}\langle\varphi_{r}|\Op{r}|\varphi_{r}\rangle\,, 
\end{eqnarray}
with $\{m^{a}_{i}\} = \{1, \dots i-1, a, i+1, \dots N\}$. We restrict ourselves to the frozen-core approximation as well as the 
single-channel approximation for the photoionization processes, allowing only
the HOMO orbital, $\varphi_{i_0}$, 
to contribute to the photoioization via direct (first order) or 
two-photon (second order) ionization probing 
different unoccupied orbitals $\varphi_a$.

Inserting the first order corrections $\alpha^{(0)}_0(t)$ and $\alpha^{a\,(0) }_i(t)$ defined in Eqs.~\eqref{eq:alpha_0.firstorder00} 
and~\eqref{eq:alpha_ia.firstorder2}  
into Eq.~\eqref{eq:a_ik.second} gives 
\begin{eqnarray}
  \label{eq:alpha_ik_sec.preliminary}
  \alpha^{\boldsymbol{k}(2)}_{i_0}(t) &=&
  -\langle\Phi^{\boldsymbol{k}}_{i_0}|\Op{r}|\Phi_0\rangle\langle\Phi_0|\Op{r}|\Phi_0\rangle\,\\
  &&\quad\quad\times\,\int^t_{-\infty}e^{i(\epsilon^k_{i_0}-\epsilon_0)t^\prime}\boldsymbol{\mathcal{E}}(t^\prime)\,\int_{-\infty}^{t^\prime}\boldsymbol{\mathcal{E}}(t^{\prime\prime})\,dt^{\prime\prime}\,dt^\prime\nonumber\\ 
  &&-\sum_{\substack{b\notin \text{occ}\\j\in\text{occ}}}\langle\Phi^{\boldsymbol{k}}_{i_0}|\Op{r}|\Phi^b_j\rangle\langle\varphi_b|\Op{r}|\varphi_j\rangle\,\int^t_{-\infty}e^{i(\epsilon^k_{i_0}-\epsilon^b_j)t^\prime}\boldsymbol{\mathcal{E}}(t^\prime)\,\nonumber\\
  &&\quad\quad\quad\quad\quad\quad\times\,  \int_{-\infty}^{t^\prime}e^{i(\epsilon^b_j-\epsilon_0)}\boldsymbol{\mathcal{E}}(t^{\prime\prime})\,dt^{\prime\prime} 
\,dt^\prime\nonumber \,.
\end{eqnarray}

\subsection{\textbf{\sffamily{Orientation-dependent momentum distribution}}}
Two steps are still required to compute the laboratory frame quantities 
$\alpha^{\boldsymbol{k}^\prime(1)}_{i_0}(t;\gamma_\mathcal{R})$
and $\alpha^{\mathbf{k}^\prime(2)}_{i_0}(t;\gamma_\mathcal{R})$. First, if $\boldsymbol{\mathcal{E}}(t) = \mathcal{E}_{\mu_0}(t)$ is known in $\mathcal{R}^\prime$, 
the component $\mathcal{E}_{\mu_0}(t)$ must be projected into the molecular frame in order to evaluate  
all tensor elements. The
orientation of $\mathcal{R}$ with respect to  $\mathcal{R}^\prime$ is defined
by the Euler angles $\gamma_{\mathcal{R}}=(\alpha,\beta,\gamma)$. The second step involves rotation 
of the photoelectron direction of emission from $\mathcal{R}$ to $\mathcal{R}^\prime$,  which is performed using the inverse of
the transformation used in step 1.
 
In $\mathcal{R}^\prime$, the electric field is decomposed into the spherical unit
vectors $\epsilon^{\prime}_{\mu}$, with $\mu=\pm 1,0$,
\begin{eqnarray*}
  \boldsymbol{\mathcal{E}}(t) = \sum_{\mu}\,\mathcal{E}_{\mu}\,\hat{\epsilon}^{\prime\,*}_{\mu} =\sum_{\mu}\,(-1)^{\mu}\mathcal{E}_{\mu}\,\hat{\epsilon}^{\prime}_{-\mu}\,,
\end{eqnarray*}
such that
 $ \boldsymbol{\mathcal{E}}(t)\cdot\hat{\epsilon}^\prime_{\mu_0} = \mathcal{E}_{\mu_0}(t)\,$. 
We recall that $\mu_0=\pm 1,0$ defines the left $(+1)$, right $(-1)$ or linear $(0)$
polarization direction of  
the field. 
The spherical unit vector $\epsilon^\prime_{\mu_0}$ 
can be written in terms of its molecular-frame counterparts, $\epsilon_{\mu}$,
\begin{eqnarray}
  \label{eq:transformation}
  \epsilon^\prime_{\mu_0} &=& \sum_{\mu=\pm1,0}\mathcal{D}^{(1)}_{\mu,\mu_0}(\gamma_{\mathcal{R}})\,\epsilon_{\mu}\,,
\end{eqnarray}
where  $\mathcal{D}^{(1)}_{\mu,\mu_0}(\gamma_{\mathcal{R}})$ are the
components of  the Wigner rotation matrix.       
Rotations between the two 
coordinate systems, $\mathcal{R}$ and $\mathcal{R}^\prime$,  are performed
following the convention
for the Wigner rotation matrices of             
Ref.~\cite{edmonds2016angular}. The 
tensor operator $\Op{r}$ is also projected into the spherical unit vectors in $\mathcal{R}$,         
\begin{eqnarray*}
  \Op{r} &=&\sum_{\mu}\,(-1)^{\mu} \Op{r}_{\mu}\,\epsilon_{-\mu}\,,
\end{eqnarray*}
such that $\Op{r}\cdot\epsilon_{\mu_0}= \Op{r}_{\mu_0}$. Finally, 
in the molecular frame, the dipole interaction reads 
\begin{eqnarray*}
  \boldsymbol{\mathcal{E}}(t)\cdot\Op{r} &=& \mathcal{E}_{\mu_0}(t)\,\sum_{\mu}\mathcal{D}^{(1)\,*}_{\mu,\mu_0}(\gamma_{\mathcal{R}})\,\Op{r}_{\mu}
\end{eqnarray*}
with $\mu=\pm1,0$ and                            
the matrix elements of the interaction become  
\begin{eqnarray}
  \label{eq:dipole.molecularframe2} \mathcal{E}_{\mu_0}(t)\langle\varphi_p|\epsilon^{\prime\,*}_{\mu_0}\cdot\Op{r}|\varphi_q\rangle &=& 
  \mathcal{E}_{\mu_0}(t)\\
  &&\times\sum_{\mu}\mathcal{D}^{(1)\,*}_{\mu,\mu_0}(\gamma_{\mathcal{R}})\,\langle\varphi_p|\Op{r}_{\mu}|\varphi_q\rangle\,,\nonumber 
\end{eqnarray}
where $\varphi_{p}$, $\varphi_{q}$ symbolize two arbitrary spin-orbitals, and 
Eq.~\eqref{eq:dipole.molecularframe2} is also
valid for  
scattering states. Here, the  
$z^\prime > 0$ axis is defined by the propagation direction of
the laser beam, normal to the $(x^\prime,y^\prime)$-plane (polarization plane
for $\mu_0 = \pm 1$). 

The second type of rotation involves projection of the direction of
photoelectron emission from the molecular to the laboratory frame. In the molecular frame, the direction of photoelectron emission is obtained by
expanding the scattering wave function into spherical harmonics, 
\begin{eqnarray}
  \label{eq:epolyexpansion}
  \varphi^-_{\boldsymbol{k}}(\boldsymbol{r}) &=&\sum_{\ell,m}\varphi^-_{k,\ell,m}(\boldsymbol{r})\,Y^{\ell}_m(\theta_{\boldsymbol{k}},\phi_{\boldsymbol{k}})\, ,
\end{eqnarray}
where $\theta_{\boldsymbol{k}}$ and $\phi_{\boldsymbol{k}}$ correspond to the
polar and azimuthal angles  
in the molecular frame. 
To obtain the  
anisotropy parameters, evaluation of the product of two spherical
harmonics, $Y^{\ell}_m(\theta_{\boldsymbol{k}},\phi_{\boldsymbol{k}})Y^{\ell^\prime}_{m^\prime}(\theta_{\boldsymbol{k}},\phi_{\boldsymbol{k}})$, is required.
This can be done   
using the well known expression~\cite{szabo2012modern}  
\begin{widetext}
\begin{eqnarray}
  \label{eq:YY.mol} 
  Y^{\ell}_m(\theta_{\boldsymbol{k}},\phi_{\boldsymbol{k}}) Y^{*\ell^\prime}_{m^\prime}(\theta_{\boldsymbol{k}},\phi_{\boldsymbol{k}}) 
  &=& (-1)^m 
\sum_{L}\tilde{\gamma}^{L}_{\ell,\ell^\prime}\, Y^{L}_{m-m^\prime}(\theta_{\boldsymbol{k}},\phi_{\boldsymbol{k}})
\begin{pmatrix} \ell & \ell^\prime & L\vspace*{0.31cm}\\ 0 & 0 & 0 \end{pmatrix} 
 \begin{pmatrix} \ell & \ell^\prime & L\vspace*{0.31cm}\\ m & -m^\prime &m^\prime-m\end{pmatrix}.
\end{eqnarray}
where 
\begin{eqnarray*}
  \tilde{\gamma}(\ell,\ell^\prime,L) & \equiv&\sqrt{\dfrac{(2L+1)}{4\pi}(2\ell+1)(2\ell^\prime+1)}\,.
\end{eqnarray*}
The last step consists in writing  the direction of photoelectron emission --- determined by
$Y^{L}_{m-m^\prime}(\theta_{\boldsymbol{k}},\phi_{\boldsymbol{k}})$ appearing in the rhs of Eq.~\eqref{eq:YY.mol} --- with respect to the
laboratory frame. This is achieved using the inverse of the 
transformation
~\eqref{eq:transformation}, 
\begin{eqnarray}
  \label{eq:YY.lab} Y^{L}_{m-m^\prime}(\theta_{\boldsymbol{k}},\phi_{\boldsymbol{k}})&=&\mathcal{D}^{-1}Y^{L}_{m-m^\prime}(\theta^\prime,\phi^\prime) 
  =\sum_{M^\prime}\mathcal{D}^{(L)\dagger}_{M,m-m^\prime}(\gamma_{R})\,Y^{L}_{M}(\theta_{\boldsymbol{k}^\prime} ,\phi_{\boldsymbol{k}^\prime})
\nonumber\\                &=&  
\sum_{M^\prime}(-1)^{m-m^\prime-M}\mathcal{D}^{L}_{m^\prime-m,-M}(\gamma_{\mathcal{R}})
\,Y^{L}_{M}(\theta_{\boldsymbol{k}^\prime},\phi_{\boldsymbol{k}^\prime})\,, 
\end{eqnarray}
\end{widetext}
which allows us to write  Eq.~\eqref{eq:YY.mol} as a function of  the polar and 
azimuthal angles  
$\theta_{\boldsymbol{k}^\prime}$ and
$\phi_{\boldsymbol{k}^\prime}$  
in the laboratory frame.  The 
reason for  
rotating the product of two spherical harmonics instead of
evaluating the product of two rotated harmonics is two-fold: first, it 
results in 
a more compact expression 
of the anisotropy parameters, by writing the final expression as a product  of
three instead of four 
Wigner rotation matrices,  
and secondly, it 
allows for  
a more straightforward analysis of the properties of the Wigner $3j$ symbols 
appearing in  
the anisotropy parameters under polarization reversal 
which is of
interest when evaluating the PECD.   

Finally, applying  
Eq.~\eqref{eq:YY.lab} to the spherical harmonics in the
rhs of Eq.~\eqref{eq:YY.mol} gives 
\begin{widetext}
\begin{eqnarray}
  \label{eq:YY.mol.to.YY.lab0} 
  Y^{\ell}_m(\theta_{\boldsymbol{k}},\phi_{\boldsymbol{k}}) Y^{*\ell^\prime}_{m^\prime}(\theta_{\boldsymbol{k}},\phi_{\boldsymbol{k}}) 
  &=& (-1)^{m^\prime}\sum_{L,M}\tilde{\gamma}^L_{\ell,\ell^\prime}  
 \begin{pmatrix} \ell & \ell^\prime & L\vspace*{0.31cm}\\ 0 & 0 & 0 \end{pmatrix} 
 \begin{pmatrix} \ell & \ell^\prime & L\vspace*{0.31cm}\\ m & -m^\prime &m^\prime-m\end{pmatrix} (-1)^{-M}\mathcal{D}^{(L)}_{m^\prime-m,-M}(\gamma_{\mathcal{R}})\,Y^{L}_M(\theta_{\boldsymbol{k}^\prime},\phi_{\boldsymbol{k}^\prime} )\nonumber
\end{eqnarray}
or, equivalently,
\begin{eqnarray}
  \label{eq:YY.mol.to.YY.lab} 
  Y^{\ell}_m(\theta_{\boldsymbol{k}},\phi_{\boldsymbol{k}}) Y^{*\ell^\prime}_{m^\prime}(\theta_{\boldsymbol{k}},\phi_{\boldsymbol{k}}) 
  &=& (-1)^{m^\prime}\sum_{L,M}\zeta^{L,M}_{\ell,\ell^\prime}  
                    \begin{pmatrix} \ell & \ell^\prime & L\vspace*{0.31cm}\\ 0 & 0 & 0 \end{pmatrix} 
 \begin{pmatrix} \ell & \ell^\prime & L\vspace*{0.31cm}\\ m & -m^\prime &m^\prime-m\end{pmatrix}
                      \mathcal{D}^{(L)}_{m^\prime-m,-M}(\gamma_{\mathcal{R}})\,P^{M}_L(\cos\theta_{\boldsymbol{k}^\prime})\,e^{i\,M\phi_{\boldsymbol{k}^\prime}}\,,\quad\quad
\end{eqnarray}
\end{widetext}
where $P^M_L(\cdot)$ denotes the associate Legendre polynomials and 
\begin{eqnarray}
  \zeta^{L,M}_{\ell,\ell^\prime} &=& \dfrac{(2L+1)}{4\pi}\sqrt{(2\ell+1)(2\ell^\prime+1)\dfrac{(L-M)!}{(L+M)!}}\,.
\end{eqnarray}
The  
orientation-averaged laboratory-frame anisotropy parameters
associated to the one- and two-photon ionization pathways and their interference
defined in Sec.~\ref{sec:orientation_averaged} are evaluated in the following.

\section{\textbf{\sffamily{Laboratory-frame anisotropy parameters }}}
\label{sec:section2}
\subsection{\textbf{\sffamily{Anisotropy parameters $\beta^{(\mu_0)1ph}_{L,M}$}}}

Using Eq.~\eqref{eq:dipole.molecularframe2}, the molecular-frame Eq.~\eqref{eq:alpha_ik_first.preliminary}                                         
becomes 
\begin{eqnarray}
  \label{eq:ak1.rot2}
  \alpha^{\boldsymbol{k}(1)}_{i_0}(t;\gamma_{\mathcal{R}}) &=& i\sum_{\mu}\langle\Phi^{\boldsymbol{k}}_{i_0}|\Op{r}_{\mu}|\Phi_{0}\,\rangle\,\mathcal{D}^{*(1)}_{\mu,\mu_0}(\gamma_{\mathcal{R}})\\
                                                       && \quad\quad\quad\quad\times\,\int_{-\infty}^{t}e^{-i(\epsilon^k_{i_0}-\epsilon_0)}\mathcal{E}_{\mu_0}(t^\prime)\,dt^\prime\,.\nonumber
\end{eqnarray}
Inserting  
the partial wave expansion
~\eqref{eq:epolyexpansion}, we find 
\begin{eqnarray}
  \label{eq:aik.lab}
  \alpha^{\boldsymbol{k}(1)}_{i_0}(t;\gamma_{\mathcal{R}}) &=& i\sum_{\mu}\sum_{\ell,m}\mathcal{M}^{(1)}_{k,\ell,m}\,\mathcal{D}^{*(1)}_{\mu,\mu_0}(\gamma_{\mathcal{R}})\\\nonumber                            &&\quad\,\times\,Y^{\ell}_{m}(\theta_{\boldsymbol{k}},\phi_{\boldsymbol{k}})\int_{-\infty}^{t}e^{-i(\epsilon^k_{i_0}-\epsilon_0)}\mathcal{E}_{\mu_0}(t^\prime)\,dt^\prime \,,
\end{eqnarray}
where $\mathcal{M}^{\mu}_{k,\ell,m}\equiv \langle\Phi^{k,\ell,m}_{i_0}|\Op{r}_{\mu}|\Phi_0\rangle$ is the partial wave decomposition of the matrix elements
in Eq.~\eqref{eq:ak1.rot2}. Making use of Eq.~\eqref{eq:YY.mol.to.YY.lab} and  evaluating $\alpha^{\boldsymbol{k}}_{i_0}(t;\gamma_\mathcal{R}) \alpha^{*\boldsymbol{k}}_{i_0}(t;\gamma_\mathcal{R})$
results in 
\begin{widetext}
\begin{eqnarray}
  \label{eq:beta1.gamma} \alpha^{\boldsymbol{k}^\prime(1)}_{i_0}(\gamma_{\mathcal{R}})\alpha^{*\boldsymbol{k}^\prime(1)}_{i_0}(\gamma_{\mathcal{R}})&=&(-1)^{-\mu_0}
  \sum_{\mu,\mu^\prime}\sum_{\substack{\ell,m\\\ell^\prime,m^\prime}}
  \mathcal{M}^{\mu}_{k,\ell,m} \mathcal{M}^{*\mu^\prime}_{k,\ell^\prime,m^\prime}    \int^{t}_{-\infty}e^{i(\epsilon^k_{i_0}-\epsilon_0)t^\prime}\mathcal{E}_{\mu_0}(t^\prime)\,dt^\prime \int^{t}_{-\infty}e^{-i(\epsilon^k_{i_0}-\epsilon_0)t^\prime}\mathcal{E}^{*}_{\mu_0}(t^\prime)\,dt^\prime\nonumber\\
  &&\times\sum_{L,M}\tilde{\gamma}^{L,M}_{\ell,\ell^\prime}\,(-1)^{-m^\prime+\mu}
 \begin{pmatrix} \ell & \ell^\prime & L\vspace*{0.31cm}\\ 0 & 0 & 0 \end{pmatrix} 
 \begin{pmatrix} \ell & \ell^\prime & L\vspace*{0.31cm}\\ m & -m^\prime &m^\prime-m\end{pmatrix} \,P^{M}_{L}(\cos\theta_{\boldsymbol{k}^\prime})\,e^{iM\phi_{\boldsymbol{k}^\prime}}\nonumber\\[0.3cm] 
&& \times\mathcal{D}^{(1)}_{-\mu,-\mu_0}(\gamma_{\mathcal{R}})
\mathcal{D}^{(1)}_{\mu^\prime,\mu_0}(\gamma_{\mathcal{R}})\mathcal{D}^{(L)}_{m^\prime-m,-M}(\gamma_{\mathcal{R}})\,. 
\end{eqnarray}
\end{widetext}
We recall that $\mu_0=\pm 1,0$ defines the light polarization direction in the laboratory frame.
Following Eq.~\eqref{eq:1phcontr}, integrating Eq.~\eqref{eq:beta1.gamma} over
the Euler angles leads to the contribution of the one-photon pathway to the
momentum distribution. Integrating a  product of three Wigner $3j-$ symbols 
over the Euler angles can be evaluated according to~\cite{edmonds2016angular}
\begin{eqnarray}
  \label{eq:3j.int}
&& \int\mathcal{D}^{(\ell_1)}_{m_1,m^\prime_1}(\gamma_{\mathcal{R}}) \mathcal{D}^{(\ell_2)}_{m_2,m^\prime_2}(\gamma_{\mathcal{R}})\mathcal{D}^{(\ell_3)}_{m_3,m^\prime_3}(\gamma_{\mathcal{R}}) \dfrac{d^3\gamma_{\mathcal{R}}}{8\pi^2}\nonumber\\[0.2cm] 
         &&\quad\quad\quad=\begin{pmatrix} \ell_1 & \ell_2 & \ell_3\vspace*{0.31cm}\\ m_1 & m_2 & m_3\end{pmatrix} 
\begin{pmatrix} \ell_1 & \ell_2 & \ell_3\vspace*{0.31cm}\\ m^\prime_1 & m^\prime_2 & m^\prime_3\end{pmatrix}\,.\\[0.2cm]\nonumber 
\end{eqnarray}
Another useful property concerns the product of two
Wigner rotation matrices given by~\cite{edmonds2016angular}
\begin{eqnarray}
\label{subeq:property.product}
&&\mathcal{D}^{(\ell_1)}_{m_1,m^\prime_1}(\gamma_{\mathcal{R}})
\mathcal{D}^{(\ell_2)}_{m_2,m^\prime_2}(\gamma_{\mathcal{R}})
 =\sum_{\ell}
 (2\ell+1)\,\mathcal{D}^{*(j)}_{-m_{12},-m^\prime_{12}}(\gamma_{\mathcal{R}})
        \nonumber\\
    &&\quad\quad\quad\quad\times\,
       \begin{pmatrix} \ell_1 & \ell_2 & \ell \vspace*{0.33cm} \\ 
        m_1 & m_2 & -m_{12} 
       \end{pmatrix}
       \begin{pmatrix} \ell_1 & \ell_2 & \ell \vspace*{0.33cm} \\ 
       m^\prime_1 & m^\prime_2 & -m^\prime_{12} 
       \end{pmatrix}\,,
\end{eqnarray}
with $m_{12}=m_1+m_2$ and $m^\prime_{12}=m^\prime_1+m^\prime_2$. Upon integration 
of Eq.~\eqref{eq:beta1.gamma} over the Euler angles using Eq.~\eqref{eq:3j.int} 
and        
equating the result                    
with Eq.~\eqref{eq:1phcontr}, 
the contribution from the one-photon ionization pathway to the orientation-averaged anisotropy parameters is obtained, 
\begin{widetext}
\begin{eqnarray}
  \label{eq:Beta1ph.final}
  \beta^{(\mu_0)1ph}_{L,M}(\epsilon_k) &=&2\pi\,(2L+1)\sqrt{(2\ell+1)(2\ell^\prime+1)}\,(-1)^{-\mu_0}\\ 
                                       && \times\sum_{\mu\,\mu^\prime}\sum_{\substack{\ell,m\\\ell^\prime,m^\prime}}
\mathcal{M}^{\mu}_{k,\ell,m} \mathcal{M}^{*\mu^\prime}_{k,\ell^\prime,m^\prime}
 \int^{t}_{-\infty}e^{i(\epsilon^k_{i_0}-\epsilon_0)t^\prime}\mathcal{E}_{\mu_0}(t^\prime)\,dt^\prime
 \int^{t}_{-\infty}e^{-i(\epsilon^k_{i_0}-\epsilon_0)t^\prime}\mathcal{E}^{*}_{\mu_0}(t^\prime)\,dt^\prime\nonumber\\
 &&\times(-1)^{-m^\prime+\mu}
\begin{pmatrix} \ell & \ell^\prime & L\vspace*{0.31cm}\\ 0 & 0 & 0 \end{pmatrix} 
\begin{pmatrix} \ell & \ell^\prime & L\vspace*{0.31cm}\\ m & -m^\prime &m^\prime-m\end{pmatrix}
\begin{pmatrix} 1 & 1 & L\vspace*{0.31cm}\\ -\mu & \mu^\prime & m^\prime-m\end{pmatrix} 
\begin{pmatrix} 1 & 1 & L\vspace*{0.31cm}\\ -\mu_0 & \mu_0 & 0\end{pmatrix}\,\delta_{M,0}\,.\nonumber 
\end{eqnarray}
\end{widetext}

Next, we analyze the symmetry properties of $\beta^{(\mu_0)1ph}_{\ell,m}$ under
polarization reversal, i.e, $\mu_0\rightarrow -\mu_0$, which is of interest
in view of defining the PECD in Sec.~\ref{sec:section3}.
We start by considering the  fourth $3j$ Wigner symbol in Eq.~\eqref{eq:Beta1ph.final},
which is the only polarization dependent term in Eq.~\eqref{eq:Beta1ph.final},  and denote it by $\mathcal{S}_{1ph}(\mu_0)$.
Under polarization reversal, $\mu_0\rightarrow -\mu_0$,  $\beta^{(\mu_0)1ph}_{L,0}$
transforms to $\beta^{(-\mu_0)1ph}_{L,0}$, and 
\begin{eqnarray}
  \label{eq:S1ph}
  \mathcal{S}_{1ph}(-\mu_0)&=& 
\begin{pmatrix} 1 & 1 & L\vspace*{0.31cm}\\ \mu_0 & -\mu_0 & 0\end{pmatrix} 
=(-1)^{2+L}\begin{pmatrix} 1 & 1 & L\vspace*{0.31cm}\\ -\mu_0 & \mu_0 & 0\end{pmatrix}\nonumber\\
                            &=& (-1)^{L}\,\mathcal{S}(+\mu_0)\,,
\end{eqnarray}
due to 
the symmetry property for the Wigner $3j$ symbols, 
\begin{eqnarray}
  \label{eq:wigner.property.arrow}
\begin{pmatrix} j_1 & j_2 & j\vspace*{0.31cm}\\ m_1& m_2 & m \end{pmatrix}
                    &=& (-1)^{j_1+j_2+j}
\begin{pmatrix} j_1 & j_2 & j\vspace*{0.31cm}\\ -m_1& -m_2 & -m
\end{pmatrix}\,.\quad\quad
\end{eqnarray}
Since no other term depends on $\mu_0$, Eq.~\eqref{eq:S1ph} implies 
\begin{eqnarray}
  \label{eq:betaexc1ph}
  \beta^{(+\mu_0)1ph}_{L,0} &=&(-1)^L\, \beta^{(-\mu_0)1ph}_{L,0}\,.
\end{eqnarray}
For $\mu_0 = 0$ (linear polarization),            
$\mathcal{S}(\mu_0=0)$ allows for only 
even numbers $1+1+L$ to be non-zero. Therefore $\beta^{(0)1ph}_{L=1,M=0}=0$. Furthermore, for
$\mu_0=\pm 1$,  $L=1$ is the only contributing term (up to second order), as it should be,  since both the third and fourth
Wigner $3j$ symbol in Eq.~\eqref{eq:Beta1ph.final} are non-zero only for 
$0\le L \le 2$.  
This implies $\beta^{(\mu_0)1ph}_{3,0} =0$. 
Finally, using Eq.~\eqref{eq:betaexc1ph}, Eq.~\eqref{eq:1phcontr} becomes, under the 
exchange $\mu_0\rightarrow -\mu_0$,
  \begin{eqnarray}
  \label{eq:ForPecd1ph_a}
    \dfrac{d^2\sigma^{(-\mu_0)1ph}}{d\epsilon_k\,d\Omega_{\boldsymbol{k}^\prime}}&=&
  \sum^2_{L=0}\beta^{(-\mu_0)1ph}_{L,0}P^{0}_{L}(\cos\theta_{\boldsymbol{k}^\prime})\nonumber\\
 &=& \sum^2_{L=0}(-1)^L\beta^{(\mu_0)1ph}_{L,0}P^{0}_{L}(\cos\theta_{\boldsymbol{k}^\prime})\,.
\end{eqnarray}

\subsection{\textbf{\sffamily{Anisotropy parameters $\beta^{(\mu_0)2ph}_{L,M}$}}}

Rotating all tensor elements of the dipole interaction in Eq.~\eqref{eq:alpha_ik_sec.preliminary} into the molecular frame leads to
\begin{widetext}
\begin{eqnarray}
  \label{eq:Ak2}
  \alpha^{\boldsymbol{k}(2)}_{i_0}(t;\gamma_{\mathcal{R}}) &=&
  -\sum_{\mu,\mu^\prime}\mathcal{D}^{*(1)}_{\mu,\mu_0}(\gamma_\mathcal{R})\mathcal{D}^{*(1)}_{\mu^\prime,\mu_0}(\gamma_\mathcal{R})\Big[ \langle\Phi^{\boldsymbol{k}}_{i_0}|\Op{r}_{\mu}|\Phi_0\rangle\langle\Phi_0|\Op{r}_{\mu^\prime}|\Phi_0\rangle\,\int^t_{-\infty}e^{i(\epsilon^k_{i_0}-\epsilon_0)t^\prime}\mathcal{E}_{\mu_0}(t^\prime)\,\int_{-\infty}^{t^\prime}\mathcal{E}_{\mu_0}(t^{\prime\prime})\,dt^{\prime\prime}\,dt^\prime\\\nonumber
    &&\quad\quad\quad\quad\quad\quad\quad\quad+\,\sum_{\substack{b\notin \text{occ}\\j\in\text{occ}}}\langle\Phi^{\boldsymbol{k}}_{i_0}|\Op{r}_{\mu}|\Phi^b_j\rangle\langle\Phi^b_j|\Op{r}_{\mu^\prime}|\Phi_0\rangle\int^t_{-\infty}e^{i(\epsilon^k_{i_0}-\epsilon^b_{i_0})t^\prime}\mathcal{E}_{\mu_0}(t^\prime)\,
  \int_{-\infty}^{t^\prime}e^{i(\epsilon^b_{i_0}-\epsilon_0)}\mathcal{E}_{\mu^\prime_0}(t^{\prime\prime})\,dt^{\prime\prime}\,dt^\prime\,\Big].\\\nonumber 
\end{eqnarray}
In order to write the final expression as a product of three Wigner rotation
matrices --- for easy integration over the Euler angles --- it is convenient to
apply Eq.~\eqref{subeq:property.product} to the two Wigner rotation matrices in
Eq.~\eqref{eq:Ak2}. 
Defining two-photon tensor matrix elements, 
\begin{eqnarray}
  T^{i_0,r}_{\mu,\mu^\prime}(k,l,m) =
  \begin{dcases} 
    \langle\varphi^-_{k,\ell,m}|\Op{r}_{\mu}|\varphi_{i_0}\rangle\langle\Phi_0|\Op{r}_{\mu^\prime}|\Phi_0\rangle\,, &\text{if\, $r = i_0$} \\[0.4cm]\nonumber
    \langle\varphi^-_{k,\ell,m}|\Op{r}_{\mu}|\varphi_{r}\rangle\langle\varphi_r|\Op{r}_{\mu^\prime}|\varphi_{i_0}\rangle\,, &\text{if\, $r > i_0$} \,,
   \end{dcases}
\end{eqnarray}
together with the control-dependent quantity 
\begin{eqnarray}
  \zeta^{i_0,r}_{\mu_0}(t;k) =
  \begin{dcases} 
    \int^{t}_{-\infty}e^{(\epsilon^k_{i_0}-\epsilon_0)t^\prime}\mathcal{E}_{\mu_0}(t^\prime)\int^{t^\prime}_{-\infty}\mathcal{E}_{\mu_0}(t^{\prime\prime})\,dt^{\prime\prime}\,dt^\prime\,, &\text{if\, $r = i_0$} \\[0.4cm]
    \int^{t}_{-\infty}e^{(\epsilon^k_{i_0}-\epsilon^r_{i_0})t^\prime}\mathcal{E}_{\mu_0}(t^\prime)\int^{t^\prime}_{-\infty}e^{i(\epsilon^r_{i_0} - \epsilon_{0})t^{\prime\prime}}\mathcal{E}_{\mu_0}(t^{\prime\prime})\,dt^{\prime\prime}\,dt^\prime\,, &\text{if\, $r > i_0$}\,,
   \end{dcases}
\end{eqnarray}
Eq.~\eqref{eq:Ak2} becomes
\begin{eqnarray}
  \label{eq:Ak2.final2}
  \alpha^{\boldsymbol{k}(2)}_{i_0}(t;\gamma_{\mathcal{R}}) &=&
  -\sum_{\mu,\mu^\prime}\sum_{\ell,m}\sum_{r\geqslant i_0}T^{i_0,r}_{\mu,\mu^\prime}(k,\ell,m)\,\zeta^{i_0,r}_{\mu_0}(t;k) 
  \sum^{2}_{Q_1=0}g^{(Q_1)}_{\mu,\mu^\prime}(\mu_0)\,\mathcal{D}^{(Q_1)}_{-\mu-\mu^\prime,0}(\gamma_{\mathcal{R}})\, Y^{\ell}_{m}(\Omega_{\boldsymbol{k}})
\end{eqnarray}
\end{widetext}
where 
\begin{eqnarray}
  \label{eq.gQ.def}
  g^{(Q_1)}_{\mu,\mu^\prime}(\mu_0) &\equiv& (2Q_1+1)
 \begin{pmatrix} 1 & 1 & Q_1\vspace*{0.31cm}\\ \mu & \mu^\prime & -\mu^\prime-\mu \end{pmatrix} 
 \begin{pmatrix} 1 & 1 & Q_1\vspace*{0.31cm}\\ \mu_0 & \mu_0
                   & -2\mu_0 \end{pmatrix}\,.\nonumber
\end{eqnarray}

Analogously to the first-order correction, multiplication of Eq.~\eqref{eq:Ak2.final2} 
with its complex conjugate, followed by rewriting  
the product
$Y^\ell_{m}(\Omega_{\boldsymbol{k}})Y^{\ell^\prime\,*}_{m^\prime}(\Omega_{\boldsymbol{k}})$ using Eq.~\eqref{eq:YY.mol.to.YY.lab}, 
we find
\begin{widetext}
\begin{eqnarray}
 \label{eq:a2.squared}
\big|\alpha^{\boldsymbol{k^\prime}(2)}_{i_0}(t;\gamma_{\mathcal{R}})\big|^2&=&
 \sum_{\mu,\mu^\prime}\sum_{\ell,m}\sum_{r\geqslant i_0}T^{i_0,r}_{\mu,\mu^\prime}(k,\ell,m)\,\zeta^{i_0,r}_{\mu_0}(t;k)\sum^{2}_{Q_1=0}g^{(Q_1)}_{\mu,\mu^\prime}(\mu_0)\nonumber\\
 &&\times \sum_{\nu,\nu^\prime}\sum_{\ell^\prime,m^\prime}\sum_{r^\prime\geqslant i_0}T^{i_0,r^\prime}_{\nu,\nu^\prime}(k,\ell^\prime,m^\prime)\,\zeta^{i_0,r^\prime}_{\mu_0}(t;k)\sum^{2}_{Q_2=0}g^{(Q_2)}_{\nu,\nu^\prime}(\mu_0)\nonumber\\
 &&\times(-1)^{-m^\prime-\nu^\prime-\nu}\,\sqrt{(2\ell+1)(2\ell^\prime+1)}\,\sum_{L,M}
 \begin{pmatrix} \ell & \ell^\prime & L\vspace*{0.31cm}\\ 0 & 0 & 0 \end{pmatrix}\, 
 \begin{pmatrix} \ell & \ell^\prime & L\vspace*{0.31cm}\\ m& -m^\prime  & m^\prime-m \end{pmatrix}\,\frac{(2L+1)}{4\pi}\sqrt{\dfrac{(L-M)!}{(L+M)!}}\,\nonumber\\[0.3cm] 
                      &&\times
\mathcal{D}^{(Q_1)}_{-\mu^\prime-\mu,-2\mu_0}(\gamma_{\mathcal{R}})\,
\mathcal{D}^{(Q_2)}_{\nu^\prime+\nu,2\mu_0}(\gamma_{\mathcal{R}})\,
\mathcal{D}^{(L)}_{m^\prime-m;-M}(\gamma_{\mathcal{R}})\,P^{M}_{L}(\cos\theta_{\boldsymbol{k}^\prime})\,e^{iM\phi_{\boldsymbol{k}^\prime}}.
\end{eqnarray}
Using~Eq.~\eqref{eq:3j.int}, integration of Eq.~\eqref{eq:a2.squared} over the
Euler angles gives the anisotropy parameters $\beta^{(\mu_0)2ph}_{L,M}$, associated with the two-photon
ionization process,    
\begin{eqnarray}
  \label{eq:Beta2ph.last}
  \beta^{(\mu_0)2ph}_{L,M}(\epsilon_k) =(2\pi)(2L+1)&& 
  \,\,\,\sum_{\mu,\mu^\prime}\sum_{\ell,m}\sum_{r\geqslant i_0}T^{i_0,r}_{\mu,\mu^\prime}(k,\ell,m)\,\zeta^{i_0,r}_{\mu_0}(t;k)\sum^{2}_{Q_1=0}g^{(Q_1)}_{\mu,\mu^\prime}(\mu_0)\nonumber\\
  &&\times \sum_{\nu,\nu^\prime}\sum_{\ell^\prime,m^\prime}\sum_{r^\prime\geqslant i_0}T^{*i_0,r^\prime}_{\nu,\nu^\prime}(k,\ell^\prime,m^\prime)\,\zeta^{*i_0,r^\prime}_{\mu_0}(t;k)\sum^{2}_{Q_2=0}g^{(Q_2)}_{\nu,\nu^\prime}(\mu_0)\nonumber\\[0.25cm]
 &&\times\,(-1)^{-m^\prime-\nu^\prime-\nu}\,\sqrt{(2\ell+1)(2\ell^\prime+1)}\nonumber\\[0.3cm]
  &&\times \begin{pmatrix} \ell & \ell^\prime & L\vspace*{0.31cm}\\ 0 & 0 & 0 \end{pmatrix}\, 
 \begin{pmatrix} \ell & \ell^\prime & L\vspace*{0.31cm}\\ m& -m^\prime  & m^\prime-m \end{pmatrix} 
 \begin{pmatrix} Q_1 & Q_2 & L\vspace*{0.31cm}\\ -\mu^\prime-\mu & \nu^\prime+\nu & m^\prime-m \end{pmatrix}\, 
 \begin{pmatrix} Q_1 & Q_2 & L\vspace*{0.31cm}\\ -2\mu_0 & 2\mu_0 & -M \end{pmatrix}\,\delta_{M,0}\quad\quad 
\end{eqnarray}
\end{widetext}
with $\mu_0=\pm 1,0$. Since the last  Wigner $3j$ symbol in
Eq.~\eqref{eq:Beta2ph.last} is non-zero only if           
$-2\mu_0+2\mu_0 - M = 0$, 
non-vanishing Legendre coefficients are possible only for $M=0$. This translates
into a symmetry of the photoelectron probability distribution around the $z^\prime$
axis.

We now evaluate the behavior of $\beta^{(\mu_0)2ph}_{L,0}$ under
helicity exchange $\mu_0\rightarrow -\mu_0$. To this end, we 
define the fourth Wigner $3j$ symbol in Eq.~\eqref{eq:Beta2ph.last} by
$\mathcal{S}_{2ph}(\mu_0)$. Under helicity reversal, it transforms according to 
\begin{eqnarray}
  \label{eq:S2ph.minus}
  \mathcal{S}_{2ph}(-\mu_0)&=& 
\begin{pmatrix} Q_1 & Q_2 & L\vspace*{0.31cm}\\ +2\mu_0 & -2\mu_0 & 0\end{pmatrix}\nonumber\\
                    &=&(-1)^{Q_1+Q_2+L}
\begin{pmatrix} Q_1 & Q_2 & L\vspace*{0.31cm}\\ -2\mu_0 & +2\mu_0 & 0\end{pmatrix}\nonumber\\
&=&(-1)^{Q_1+Q_2+L}\,\mathcal{S}_{2ph}(+\mu_0)\,.
\end{eqnarray}
Additional $\mu_0-$dependent quantities in Eq.~\eqref{eq:Beta2ph.last} 
are $g^{(Q_1)}_{\mu,\mu^\prime}(\mu_0)$ and $g^{(Q_2)}_{\nu,\nu^\prime}(\mu_0)$,
both transforming according to
  \begin{eqnarray}
  \label{eq:obvious}
  g^{(Q_1)}_{\mu,\mu^\prime}(-\mu_0) &=& (-1)^{Q_1}\,g^{(Q_1)}_{\mu,\mu^\prime}(\mu_0)\,\nonumber\\ 
  g^{(Q_2)}_{\nu,\nu^\prime}(-\mu_0) &=& (-1)^{Q_2}\,g^{(Q_2)}_{\mu,\mu^\prime}(\mu_0)\,.
\end{eqnarray}
We thus find
\begin{eqnarray*}
  \mathcal{S}_{2ph}(-\mu_0)\,g^{(Q_1)}_{\mu,\mu^\prime}(-\mu_0)\, g^{(Q_2)}_{\nu,\nu^\prime}(-\mu_0)
  = \quad\quad\quad \quad\quad\\(-1)^L
  \mathcal{S}_{2ph}(+\mu_0)
  \times g^{(Q_1)}_{\mu,\mu^\prime}(+\mu_0)
  \times g^{(Q_2)}_{\nu,\nu^\prime}(+\mu_0)\,,
\end{eqnarray*}
which implies  
\begin{eqnarray}
  \beta^{(-\mu_0)\text{2ph}  }_{L,M}(\epsilon_{k}) & =&  (-1)^L\,\beta^{(+\mu_0)\text{2ph}}_{L,M}(\epsilon_{k})\,.
\end{eqnarray}
Analogously to $\beta^{(\mu_0)1ph}_{L,0}$ and as expected, 
$\beta^{(\mu_0)2ph}_{L,0}$ also 
changes sign only for odd $L$ and remains unchanged for even $L$ when
$\mu_0\rightarrow -\mu_0$.  However, in contrast to  $\beta^{(\mu_0)1ph}_{L,0}$, 
for which the only odd
contributing order was found to be $L=1$,  both $L=1$ and $L=3$ 
are allowed for $\beta^{(\mu_0)2ph}_{L,M}$. In fact,  since $|Q_1-Q_2|\le L\le Q_1+Q_2 $, 
cf. the  fourth Wigner $3j$ symbol in Eq.~\eqref{eq:Beta2ph.last}, and because 
$0 \le Q_1\le 2$ and $0 \le Q_2\le 2$,  the possible values for $L$ are $0 \le L\le 4$.
Finally, under polarization reversal, Eq.~\eqref{eq:2phcontr} becomes
  \begin{eqnarray}
  \label{eq:ForPecd2ph_a}
\dfrac{d^2\sigma^{(-\mu_0)2ph}}{d\epsilon_k\,d\Omega_{\boldsymbol{k}^\prime}}&=&
  \sum^4_{L=0}\beta^{(-\mu_0)2ph}_{L,0}(\epsilon_k)P^{0}_{L}(\cos\theta_{\boldsymbol{k}^\prime})\\\nonumber
  &=& \sum^4_{L=0} (-1)^L\, \beta^{(+\mu_0)2ph}_{L,0}(\epsilon_k)P^{0}_{L}(\cos\theta_{\boldsymbol{k}^\prime})\,.
\end{eqnarray}

\subsection{\textbf{\sffamily{Anisotropy parameters $\beta^{(\mu_0)int}_{L,M}$}}}
\label{sec:beta1ph.sec}

Using Eq.~\eqref{eq:aik.lab} and Eq.~\eqref{eq:Ak2.final2}
for the first and second order corrections, respectively, we obtain, after
rotation of the product $Y^{\ell}_m(\Omega_{\boldsymbol{k}})Y^{\ell^\prime}_{m^\prime}(\Omega_{\boldsymbol{k}})$
from the molecular to the laboratory frame,
according to Eq.~\eqref{eq:YY.mol.to.YY.lab},
\begin{widetext}
\begin{eqnarray}
  \label{eq:P1.1}
\alpha^{\boldsymbol{k}^\prime(1)}_{i_0}(t;\gamma_{\mathcal{R}})\,\alpha^{*\boldsymbol{k}^\prime(2)}_{i_0}(t;\gamma_{\mathcal{R}}) &=&
  -\dfrac{i}{4\pi}\,(-1)^{-\mu_0}\,\sum_{\mu}\sum_{\ell,m}\mathcal{M}^{\mu}_{k,\ell,m}\,\int_{-\infty}^{t}e^{-i(\epsilon^k_{i_0}-\epsilon_0)}\mathcal{E}_{\mu_0}(t^\prime)\,dt^\prime\\ 
  &&\times\,\sum_{\nu,\nu^\prime}\sum_{\ell^\prime,m^\prime}\sum_{r\ge i_0}\sum^{2}_{Q=0}T^{*i_0,r}_{\nu,\nu^\prime}(k,\ell^\prime,m^\prime)\,\zeta^{*i_0,r}_{\mu_0}(t;k)\,g^{(Q)}_{\nu,\nu^\prime}(\mu_0)\nonumber\\
&&
  \times\,(-1)^{-m^\prime+\mu-\nu^\prime-\nu}\,\sum^{\ell+\ell^\prime}_{L=|\ell-\ell^\prime|}
  (2L+1)\,\sqrt{(2\ell+1)(2\ell^\prime +1)}
 \begin{pmatrix} \ell & \ell^\prime & L\vspace*{0.31cm}\\ 0 & 0 & 0 \end{pmatrix} 
 \begin{pmatrix} \ell & \ell^\prime & L\vspace*{0.31cm}\\ m & -m^\prime & m^\prime-m \end{pmatrix}\nonumber\\
                      &&
  \times\,\sum^{L}_{M=-L}\sqrt{\dfrac{(L-M)!}{(L+M)!}}\,
  \Big[
   \mathcal{D}^{(1)}_{-\mu,-\mu_0}(\gamma_{\mathcal{R}})\,
    \mathcal{D}^{(Q)}_{\nu^\prime+\nu,2\mu_0}(\gamma_{\mathcal{R}})\,
  \mathcal{D}^{(L)}_{m^\prime-m,-M}(\gamma_{\mathcal{R}})     
\Big]\,\times
P^{M}_L(\cos\theta_{\boldsymbol{k}^\prime})\,e^{iM\phi_{\boldsymbol{k}^\prime}}\nonumber\,.
\end{eqnarray}
Integrating Eq.~\eqref{eq:P1.1} over the Euler angles gives the (complex)
anisotropy parameters associated with the interference term, as defined in
Eq.~\eqref{eq:beta.int.def}.  
Orientation averaging determines
the possible values of $M$ in Eq.~\eqref{eq:12phcontr.final}. In fact, using the equality 
\begin{eqnarray}
  \label{eq:integ.interf}
  && \int\mathcal{D}^{(1)}_{-\mu,-\mu_0}(\gamma_{\mathcal{R}})\,
    \mathcal{D}^{(Q)}_{\nu^\prime+\nu,2\mu_0}(\gamma_{\mathcal{R}})\,
    \mathcal{D}^{(L)}_{m^\prime-m,-M}(\gamma_{\mathcal{R}})\,d^3{\gamma_{\mathcal{R}}}\nonumber \\ 
                                                                                       &&\quad=\,8\pi^2\,
 \begin{pmatrix} 1 & Q & L\vspace*{0.31cm}\\ -\mu & \nu^\prime+\nu & m^\prime-m \end{pmatrix} 
 \begin{pmatrix} 1 & Q & L\vspace*{0.31cm}\\ -\mu_0 & 2\mu_0 & -M
  \end{pmatrix}\,,\quad\quad 
 \end{eqnarray}
 it is apparent that the second Wigner $3j$ symbol in Eq.~\eqref{eq:integ.interf} 
is non-zero only if  
$-\mu_0+2\mu_0-M = 0$. Consequently, 
\begin{eqnarray} 
  \label{eq:interf.reduced}
\begin{pmatrix} 1 & Q & L\vspace*{0.31cm}\\ -\mu_0 & 2\mu_0 & -M \end{pmatrix} 
                   &=&
\begin{pmatrix} 1 & Q & L\vspace*{0.31cm}\\ -\mu_0 & 2\mu_0 & -\mu_0 \end{pmatrix}
  \,\delta_{-M,-\mu_0}\,\quad
\end{eqnarray}
i.e., $M=
\mu_0$. Therefore, 
Eq.~\eqref{eq:2phcontr} is reduced to 
\begin{eqnarray}
  \label{eq:sigma.interf}
  \dfrac{d^2\sigma^{(\mu_0)int}}{d\epsilon_kd\Omega_{\boldsymbol{k}^\prime}}&=&  \sum^{3}_{L=0}
  \Big[
    \beta^{(\mu_0)int}_{L,\mu_0}(\epsilon_k)\,e^{i\mu_0\phi_{\boldsymbol{k}^\prime}}
+\,\, 
    \beta^{*(\mu_0)int}_{L,\mu_0}(\epsilon_k)\,e^{-i\mu_0\phi_{\boldsymbol{k}^\prime}}
  \Big]\,
P^{\mu_0}_{L}(\cos\theta_{\boldsymbol{k}^\prime})\,,
\end{eqnarray}
From Eq.~\eqref{eq:sigma.interf}, it is apparent that for $\mu_0=\pm 1$, the 
portion of the momentum distribution due to the
interference between one-photon and two-photon ionization pathways
breaks the azimuthal symmetry 
of photoelectron emission
around the  
light propagation direction $z^\prime$. 
This is in contrast to 
the contributions from $\beta^{(\mu_0)1ph}_{L,0} $ 
and $\beta^{(\mu_0)2ph}_{L,0} $ 
which we found to be symmetric 
around $z^\prime$.  
The anisotropy parameter due to interference                                                                 
is obtained upon the integration of Eq.~\eqref{eq:P1.1} over the Euler angles using
Eq.~\eqref{eq:integ.interf} and reads
\begin{eqnarray}
  \label{eq:beta.interf.def}
  \beta^{(\mu_0)int}_{L,\mu_0}(\epsilon_k) &=& -2i\pi\,(2L+1)\,\sum_{\mu}\sum_{\ell,m}\mathcal{M}^{\mu}_{k,\ell,m}\,\int_{-\infty}^{t}e^{-i(\epsilon^k_{i_0}-\epsilon_0)}\mathcal{E}_{\mu_0}(t^\prime)\,dt^\prime
\times\,\sum_{\nu,\nu^\prime}\sum_{\ell^\prime,m^\prime}\sum_{r\ge i_0}\sum^{2}_{Q=0}T^{*i_0,r}_{\nu,\nu^\prime}(k,\ell^\prime,m^\prime)\,
\zeta^{*i_0,r}_{\mu_0}(t;k)\,   \nonumber\\
&&
  \times\,(-1)^{-m^\prime+\mu-\nu^\prime-\nu}\,\sqrt{(2\ell+1)(2\ell^\prime +1)}
 \begin{pmatrix} \ell & \ell^\prime & L\vspace*{0.31cm}\\ 0 & 0 & 0 \end{pmatrix} 
 \begin{pmatrix} \ell & \ell^\prime & L\vspace*{0.31cm}\\ m & -m^\prime & m^\prime-m \end{pmatrix}
\begin{pmatrix} 1 & Q & L\vspace*{0.31cm}\\ -\mu & \nu^\prime+\nu & m^\prime-m \end{pmatrix}\nonumber\\[0.3cm]  
                      &&
  \times (-1)^{-\mu_0}\,g^{(Q)}_{\nu,\nu^\prime}(\mu_0)\,
 \begin{pmatrix} 1 & Q & L\vspace*{0.31cm}\\ -\mu_0 & 2\mu_0 & -\mu_0 \end{pmatrix} 
\sqrt{\dfrac{(L-\mu_0)!}{(L+\mu_0)!}}\,
\end{eqnarray}
\end{widetext}
Since $Q=0,1,2$, it is apparent from the fourth Wigner $3j$ symbol in 
Eq.~\eqref{eq:beta.interf.def} that $L=4$
cannot contribute.  
Furthermore, if $\mu_0=\pm 1$, it also the $L=0$ term
vashishes, and  
the allowed values are  
$L=1,2,3$ for circularly
polarized light, which explains the upper limit in the summation over $Q$ in
Eq.~\eqref{eq:sigma.interf}.

The symmetry properties of $\beta^{(\mu_0)int}_{L,M}$ under polarization
reversal are analyzed using the same technique as for the one-photon and two-photon terms.
It is worth
pointing out, however, that evaluation of symmetry properties  of 
Eq.~\eqref{eq:sigma.interf} under polarization reversal 
also requires evaluation of the angular functions under  $\mu_0\rightarrow -\mu_0$
if $\mu_0=\pm 1$. 
As previously discussed, the term $g^{(Q)}_{\nu,\nu^\prime}(\mu_0)$ in the rhs
of Eq.~\eqref{eq:beta.interf.def} transforms according to
\begin{subequations}
  \label{eq:symprop.int}
  \begin{eqnarray}
  \label{eq:inter1}
  g^{(Q)}_{\nu,\nu^\prime}(-\mu_0) =&& (-1)^Q\, g^{(Q)}_{\nu,\nu^\prime}(+\mu_0)\,. 
\end{eqnarray}
Analogously, the fourth Wigner $3j$ symbol in
Eq.~\eqref{eq:beta.interf.def},  denoted by $\mathcal{S}_{int}(\mu_0)$, 
changes, under $\mu_0\rightarrow -\mu_0$, as 
\begin{eqnarray}
  \label{eq:inter2}
  \mathcal{S}_{int}(-\mu_0) &\equiv&
\begin{pmatrix} 1 & Q & L\vspace*{0.31cm}\\ +\mu_0 & -2\mu_0 & +\mu_0 \end{pmatrix}\nonumber\\[0.3cm] 
                  &=&(-1)^{1+Q+L}\, \begin{pmatrix} 1 & Q & L\vspace*{0.31cm}\\ +\mu_0 & -2\mu_0 & +\mu_0 \end{pmatrix}\nonumber\\[0.3cm] 
                  &=&(-1)^{1+Q+L}\,S(-\mu_0)\,. 
\end{eqnarray}
Finally, defining 
\begin{eqnarray}
  \mathcal{B}(\mu_0) &\equiv& \sqrt{\dfrac{(L-\mu_0)!}{(L+\mu_0)!}}\,,\nonumber
\end{eqnarray}
it is straightforward to show that
\begin{eqnarray}
\label{eq:inter3}
\mathcal{B}(-\mu_0) &=& \mathcal{B}(+\mu_0)\,\times\, \dfrac{(L+\mu_0)!}{(L-\mu_0)!}\,.
\end{eqnarray}
\end{subequations}
Using Eqs.~\eqref{eq:symprop.int}(a)-(c), the anisotropy parameter $\beta^{(-\mu_0)int}_{L,-\mu_0}(\epsilon)$ behaves as 
\begin{eqnarray}
 \label{eq:beta.int.exchange}
  \beta^{(-\mu_0)int}_{L,-\mu_0} &=&
  (-1)^{1+L}\,\dfrac{(L-\mu_0)!}{(L+\mu_0)!}\times\,
  \beta^{(+\mu_0)int}_{L,+\mu_0} 
\end{eqnarray}
For the  
azimuthal dependency  
of Eq.~\eqref{eq:sigma.interf}, the
transformation under helicity exchange is trivial. As for the associated Legendre polynomials $P^{\mu_0}_{L}(\cdot)$, we use the well-known property
\begin{eqnarray}
  P^{-M}_{L}(X) &= & (-1)^{-M}\dfrac{(L-M)!}{(L+M)!}\times P^{M}_{L}(X)\,, 
\end{eqnarray}
which,  
together with Eq.~\eqref{eq:beta.int.exchange}, allows us to write the transformed  Eq.~\eqref{eq:sigma.interf}, for $\mu_0=\pm 1$, 
  \begin{eqnarray}
  \dfrac{d^2\sigma^{(-\mu_0)int}}{d\epsilon_k d\Omega_{\boldsymbol{k}^\prime}}&=&
  2\sum^{3}_{L=0}(-1)^{L}\Big[\cos(\mu_0\phi_{\boldsymbol{k}^\prime})\text{Re}\big[ \beta^{(+\mu_0)int}_{L,+\mu_0}\big]\\
    &&\,\, +\, \sin(\mu_0\phi_{\boldsymbol{k}^\prime})\text{Im}\big[ \beta^{(+\mu_0)int}_{L,+\mu_0}\big]\,\Big]P^{+\mu_0}_{L}(\cos\theta_{\boldsymbol{k}^\prime})\,.\nonumber 
\end{eqnarray}

\section{\textbf{\sffamily{Photoelectron circular dichroism}}}
\label{sec:section3}

The PECD is defined as the non-vanishing component of the differential photoelectron signal 
obtained with left and right circularly polarized light, 
\begin{eqnarray}
  \label{eq:pecd.def.last}
  \text{PECD}(\epsilon_k,\theta_{\boldsymbol{k}^\prime},\phi_{\boldsymbol{k}^\prime}) &=& 
\dfrac{d^2\sigma^{(\mu_0)}}{d\epsilon_k\,d\Omega_{\boldsymbol{k}^\prime}} -                 
\dfrac{d^2\sigma^{(-\mu_0)}}{d\epsilon_k\,d\Omega_{\boldsymbol{k}^\prime}}\,.
\end{eqnarray}
Both terms on the rhs of Eq.~\eqref{eq:pecd.def.last} contain 
the contributions defined in Eq.~\eqref{eq:SecondOrderApprox_split}, i.e.  contributions 
from the one-photon and two-photon ionization pathways and their interference. Therefore,
we can analyze the PECD associated with each of these pathways. In particular,    
\begin{subequations}
\begin{eqnarray}
  \label{eq:pecd.1ph.final}
  \dfrac{d^2\sigma^{(+\mu_0)1ph}}{d\epsilon_k\,d\Omega_{\boldsymbol{k}^\prime}} - 
\dfrac{d^2\sigma^{(-\mu_0)1ph}}{d\epsilon_k\,d\Omega_{\boldsymbol{k}^\prime}} &=&
\sum_{j} 2\beta^{(\mu_0)1ph}_{2j+1,0}P^{0}_{2j+1}(\cos\theta_{\boldsymbol{k}^\prime})\nonumber\\
&=&2\beta^{(+\mu_0)1ph}_{1,0}\,P^{0}_{1}\cos(\theta_{\boldsymbol{k}^\prime})\,,\nonumber\\
\end{eqnarray}
gives the contribution of  
the one-photon ioniozation pathways to the PECD.
The contribution from the two-photon ionization pathways is obtained as
\begin{eqnarray}
  \label{eq:pecd.2ph.final}
  \dfrac{d^2\sigma^{(+\mu_0)2ph}}{d\epsilon_k\,d\Omega_{\boldsymbol{k}^\prime}} - 
\dfrac{d^2\sigma^{(-\mu_0)2ph}}{d\epsilon_k\,d\Omega_{\boldsymbol{k}^\prime}} &=&
2\beta^{(+\mu_0)2ph}_{1,0}\,P^{0}_{1}\cos(\theta_{\boldsymbol{k}^\prime})\nonumber\\
&& + \, 2\beta^{(+\mu_0)2ph}_{3,0}\,P^{0}_{3}\cos(\theta_{\boldsymbol{k}^\prime})\,.\nonumber\\
\end{eqnarray}
Finally, the contribution of       
the interference between one-photon and
two-photon ionization pathways reads
\begin{eqnarray}
  \label{pecd.interf.both}
  && \dfrac{d^2\sigma^{(+\mu_0)int}}{d\epsilon_k\,d\Omega_{\boldsymbol{k}^\prime}} - 
\dfrac{d^2\sigma^{(-\mu_0)int}}{d\epsilon_k\,d\Omega_{\boldsymbol{k}^\prime}}\\ 
&&\quad\quad\quad = 4\cos(\mu_0\phi_{\boldsymbol{k}^\prime})\sum_{j}\text{Re}\big[\beta^{(+\mu_0)int}_{2j+1,+\mu_0}\big] P^{+\mu_0}_{2j+1}(\cos\theta_{\boldsymbol{k}^\prime})\nonumber\\
&&\quad\quad\quad -\, 4\sin(\mu_0\phi_{\boldsymbol{k}^\prime})\sum_{j}\text{Im}\big[\beta^{(+\mu_0)int}_{2j,+\mu_0}\big] P^{+\mu_0}_{2j}(\cos\theta_{\boldsymbol{k}^\prime})\nonumber\,.
\end{eqnarray}
Equation~\eqref{pecd.interf.both} implies that,  
in the $(z^\prime,y^\prime)$-plane, i.e., for $\phi_{\boldsymbol{k}^\prime}=\pi/2$,
PECD due to the interference term depends on the associate Legendre polynomials of even order.  
PECD thus changes sign in the forward and backward direction
defined by  the intervals $\theta_{\boldsymbol{k}}\in [0,\pi/2]$ and $\theta_{\boldsymbol{k}}\in ]\pi/2,\pi]$, respectively. 
  Conversely, projection of Eq.~\eqref{pecd.interf.both} into the  $(z^\prime,x^\prime)$-plane leads to  
an odd-order dependency,  which does not 
change sign between forward and backward directions.  In fact, the associate 
Legendre polynomials are either even or odd according to
\begin{eqnarray}
  P^{M}_L(-X) & = & (-1)^{L+M}\,P^{M}_L(X) 
\end{eqnarray}
with $M\equiv\mu_0$ and where $X\equiv\cos(\theta_{\boldsymbol{k}^\prime})$
changes sign in the forward and backward directions. Choosing
 $\phi_{\boldsymbol{k}^\prime}=\pi/2$ and accounting for the fact that $L=2$
is the possible even order for non-vanishing
$\beta^{(+\mu_0),int}_{L,+\mu_0}$ for $\mu_0=\pm 1$ as discussed in
Section~\ref{sec:beta1ph.sec}, Eq.~\eqref{pecd.interf.both} becomes
\begin{eqnarray}
  \label{pecd.interf.L2}
  &&  \left.\dfrac{d^2\sigma^{(+1)int}}{d\epsilon_k\,d\Omega_{\boldsymbol{k}^\prime}} - 
  \dfrac{d^2\sigma^{(-1)int}}{d\epsilon_k\,d\Omega_{\boldsymbol{k}^\prime}}\right\rvert_{\phi_{\boldsymbol{k}^\prime}=\pi/2 }\quad\quad\quad\quad\quad\\[0.3cm] 
  && \quad\quad\quad\quad\quad\quad\quad\quad\quad\quad\,=\,\, 
  6\,\text{Im}\big[\beta^{(+1)int}_{2,+1}\big]\,\sin(2\theta_{\boldsymbol{k}^\prime})\,,\nonumber
\end{eqnarray}
where we have used $P^{1}_{2}(\cos\theta) = -3\cos(\theta)\sin(\theta)$ together
with the trigonometric identity $\sin(2\theta) =2\cos(\theta)\sin(\theta) $.
\end{subequations}

Finally, accounting for the contributions from one-photon and two-photon ionization pathways
and their interference, given by Eqs.~\eqref{eq:pecd.1ph.final}, \eqref{eq:pecd.2ph.final}
and~\eqref{pecd.interf.L2}, 
Eq.~\eqref{eq:pecd.def.last} becomes
\begin{widetext}
\begin{eqnarray}  
  \label{eq:FinalPecdEquation}
  \text{PECD}(\epsilon_k,\theta_{\boldsymbol{k}^\prime},\phi_{\boldsymbol{k}^\prime}=\pi/2) &=&
  2\,\big[ \beta^{(+1)1ph}_{1,0}(\epsilon_k) +\beta^{(+1)2ph}_{1,0}(\epsilon_k)\big]\,\,P^{0}_{1}(\cos\theta_{\boldsymbol{k}^\prime})\nonumber\\[0.3cm]
&& +\,2\beta^{(+1)2ph}_{1,0}(\epsilon_k)\,\,P^{0}_{3}(\cos\theta_{\boldsymbol{k}^\prime})
  +\, 6\,\text{Im}\big[\beta^{(+1)int}_{2,+1}(\epsilon_k)\big]\,\sin(2\theta_{\boldsymbol{k}^\prime})\,.
\end{eqnarray}  
\end{widetext}

\section{\textbf{\sffamily{Control of PECD}}}
\label{sec:ControlPecd}
\subsection{\textbf{\sffamily{Optimization algorithm}}}
In the following, we summarize                      
the optimization algorithm.
\begin{enumerate}[label=(\alph*)]
  \item\label{enumaa} In a first step, the driving control field, $\epsilon(t)$ is parametrized following
    Eq.~(9) in the main text, imposing restrictions on the frequency components, 
    amplitudes, etc. following the guidelines in Ref.~\cite{GoetzSpa2016}.
  
  \item\label{enumbb} The anisotropy parameters $\beta^{(+)1ph}_{L,M}(\epsilon_k)$, $\beta^{(+)2ph}_{L,M}(\epsilon_k)$, 
    and $\beta^{(+)int}_{L,M}(\epsilon_k)$  in Eq.~\eqref{eq:FinalPecdEquation}
    are obtained using Eq.~\eqref{eq:Beta1ph.final}, Eq.~\eqref{eq:Beta2ph.last}
    and Eq.~\eqref{eq:beta.interf.def}, respectively. The time-integration
    appearing in the anisotropy parameters is evaluated using the parametrized
    time-dependent control field defined in step \ref{enumaa}.

   \item The total photoelectron momentum distribution is then evaluated using
     Eq.~\eqref{eq:SecondOrderApprox_split}. The individual terms are defined in
     Eqs.~\eqref{eq:1phcontr},~\eqref{eq:2phcontr} and ~\eqref{eq:12phcontr.final}.
     The maximum value of Eq.~\eqref{eq:SecondOrderApprox_split}, $\mathcal{N}$, is used to normalize 
      Eq.~\eqref{eq:FinalPecdEquation} if its value exceeds a given threshold, 
     otherwise we set $\mathrm{PECD}=0$.

    Note that, since the PECD is normalized with respect to the peak intensity of the
     photoelectron momentum distribution, also all the anisotropy parameters
     appearing in Eqs.~\eqref{eq:1phcontr},~\eqref{eq:2phcontr} and ~\eqref{eq:12phcontr.final}
     must be evaluated according to the equations stated in step \ref{enumbb}.

\end{enumerate}

\subsection{\textbf{\sffamily{Control objective}}}
If the final photoelectron energy $\epsilon^{*}_k$ is specified 
       but the direction $\theta^{*}_{\boldsymbol{k}^\prime}$ is not (for more flexibility), the  control objective
       takes the form
       \begin{eqnarray}
         \label{eq:allangles}
         \Gamma[\epsilon] = \underset{\theta_{\boldsymbol{k}^\prime}}{\mathrm{max}}\,\Big|\mathcal{N}^{-1}\,\mathrm{PECD}(\epsilon_k=\epsilon^*_k,\theta_{\boldsymbol{k}^\prime}, \phi_{\boldsymbol{k}^\prime}=\pi/2)\Big|\,,\quad
       \end{eqnarray}
       where both $\mathrm{PECD}$ and $\mathcal{N}$ depend on the control field $\epsilon(t)$.
 
       Conversely, if the peak position in energy is not specified, the
       control objective                    
       reads 
       \begin{eqnarray}
         \Gamma^\prime[\epsilon] = \underset{\epsilon_k, \theta_{\boldsymbol{k}^\prime}}{\mathrm{max}}\,\Big|\mathcal{N}^{-1}\,\mathrm{PECD}(\epsilon_k,\theta_{\boldsymbol{k}^\prime}, \phi_{\boldsymbol{k}^\prime}=\pi/2)\Big|\,, 
       \end{eqnarray}
       which gives more flexibility than one would obtain by integrating over $\epsilon_k$ since the  anisotropy parameters (that depends on $\epsilon_k$) 
       are in this case not forced to have large magnitude at all $\epsilon_k$  (only ``the'' maximum at a particular $\epsilon^*_k$ is of
       interest).

The functional is evaluated iteratively and the pulse parameters
       (required to evaluate the time-integrals in the anisotropy parameters) are
       iteratively updated using the adaptation of the Brent's principal axis method
       to a sequential scheme detailed in ~Ref.~\cite{GoetzSpa2016}. 

\section{\textbf{\sffamily{Single- and two photon angle-integrated PECD}}}
 \label{sec:HemisphereIntegrated}

In Fig.~1(b) in the manuscript, we have reported the maximum the single- and two-photon PECD   
over all angles $\theta_{\boldsymbol{k}}$ as a function of the photoelectron energy,  
which correspond to the expression defined in Eq.~\eqref{eq:allangles}, and 
where $\epsilon^*_k$ refers to the photoelectron energy defining the horizontal axis. 

As  a complementary support, we provide here the ``hemisphere-averaged'' PECD, which corresponds
to an angle-integrated (over the forward hemisphere) 
version of the PECD presented Fig.~1(b) in the
manuscript. The forward (backward) ``hemisphere-averaged'' PECD is obtained by integrating the PECD distribution  over the forward (backward) hemisphere.
In the interest of consistency with Fig.~1(a) in the manuscript  and Eq.~\eqref{eq:FinalPecdEquation}\footnote{also found in Eq.~(8) in the manuscript} in the supplemental material,
the azimuthal direction of photoelectron emission defined by $\phi_{\boldsymbol{k}^\prime}$, is kept fixed at $\phi_{\boldsymbol{k}^\prime}=\pi/2$. 
\begin{figure}[tb]
  \centering
\includegraphics[width=0.75\linewidth]{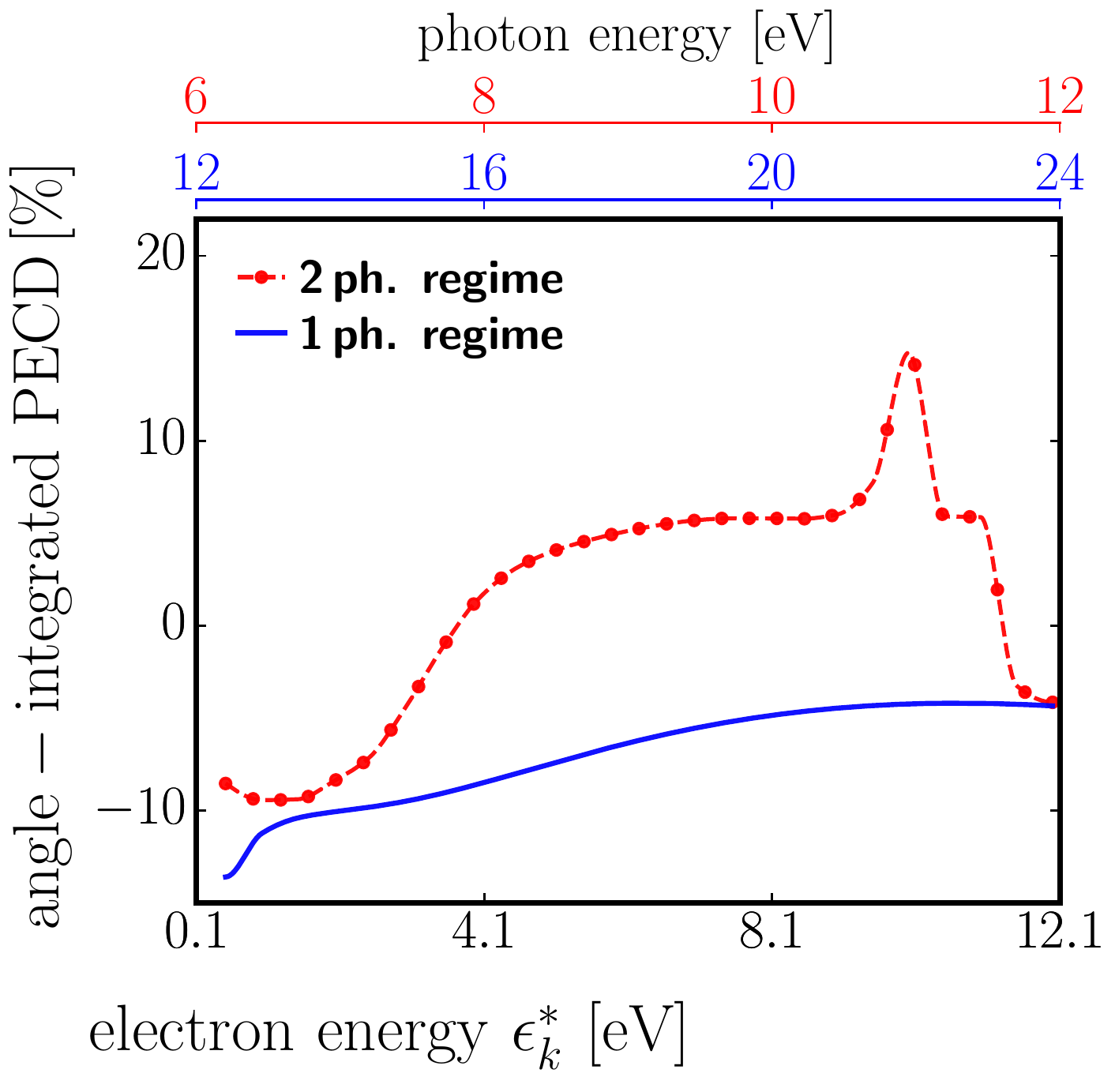}  
  \caption{Angle-averaged PECD obtained by integrating the PECD distribution in
  the forward hemisphere, defined by the light propagation direction along the
  $z$ axis, cf.~Fig.~1(a) in the manuscript.}
\label{fig:AngleIntegratedPECD}
  \end{figure}
In particular, the forward ``hemisphere-averaged'' PECD, $\rho_{f}(\epsilon_k)$, is evaluated by integrating the PECD distribution 
defined in Eq.~\eqref{eq:pecd.def.last}, using the expression 
in Eq.~\eqref{eq:FinalPecdEquation}, namely
\begin{eqnarray}
  \rho_{f}(\epsilon_k) &=&\int^{\pi/2}_{0}
  \mathrm{PECD}(\epsilon_k, \theta_{\boldsymbol{k}^\prime},\phi_{\boldsymbol{k}^\prime}=\pi/2)\,\sin\theta_{\boldsymbol{k}^\prime}\,d\theta_{\boldsymbol{k}^\prime}\,. \quad\quad
\end{eqnarray}
Conversely,  the backward ``hemisphere averaged'' counterpart, $\rho_{b}(\epsilon_k)$, is 
obtained upon integrating the PECD distribution in the interval $\theta_{\boldsymbol{k}^\prime}\in [\pi/2, \pi]$,  with the property
$\rho_{b}(\epsilon_k) =-\rho_{f}(\epsilon_k) $.

Figure~\ref{fig:AngleIntegratedPECD}
depicts the averaged PECD integrated over the forward hemisphere. For
the sake of consistency with all PECD-related quantities presented in the manuscript, the forward ``hemisphere-averaged'' PECD
shown in Fig.~\ref{fig:AngleIntegratedPECD}, 
is expressed in percentage of the peak photoelectron intensity.

\section{\textbf{\sffamily{Disentangling the optimal two-photon pathway interference scheme from
all possible contributing two-photon pathways}}}
 \label{sec:DisantanglingOptimalInterference}

Analyzing the spectrum in Fig.~3(b) in the manuscript, we have identified   all
possible two-photon pathways leading to the same final photoelectron kinetic
energy of $6.5\,$eV. The fully-optimized spectrum was found to contain specific 
frequencies components,
$\omega_1$, $\omega_2$ and $\omega_3$, which 
correspond, within a given width,  to the transition energies $\omega_{0j}$, for $j=1, \dots 4$, 
between the HOMO and $\mathrm{LUMO}+j-1$ orbitals. The field spectrum 
in Fig.~3(b) in the manuscript is characterized by a low $(\omega_1)$ and high  
$(\omega_2,\omega_3)$ frequency components. Depending on the relative delay 
between these low and high frequency components --which may be viewed as
a pump-probe scheme-- three different scenarios may be considered. The goal here
is to unravel which of these scenarios corresponds to the optimal scheme leading to
a PECD of $68\%$.

The first scenario, $\mathcal{S}_1$, consists in populating the LUMO orbital upon absorption of
a photon with energy $\omega_{10}$, which is available within the spectral width
of the low frequency component $\omega_1$. This step can
be  followed by single-photon ionization upon absorption of
a photon with energy $\omega_{02}+\delta\omega_1$ (high frequency component), which would promote the
photoelectron exactly at $6.5\,$eV. The required photon energy $\omega_{02}+\delta\omega_1$ 
is available within the spectral width of the high frequency component $\omega_2$. 
Alternatively, the photoelectron can also be promoted from the LUMO exactly at $6.5\,$eV 
upon absorption of a photon energy $\omega_{40}-\delta\omega$ (high
frequency component) where the required photon energy
$\omega_{40}-\delta\omega$ is again available within the spectral width of $\omega_4$.  
Another option for the photoelectron to acquire a kinetic energy of $6.5\,$eV
from the LUMO is to absorb a photon with energy $\omega_{50}-\delta^\prime$ (high frequency component) where again the
required photon energy is available within the spectral width of the high
frequency components of the pulse. These three possibilities defines 
the first scenario, $\mathcal{S}_1$,  and involves resonant excitation of the LUMO upon absorption of a photon energy 
within the spectral width the low energy component
$\omega_1$ followed by single-photon
ionization mediated by the high frequency components of the field. The
respective two-photon ionization scheme is  
highlighted in red in Fig.~2(f) in the manuscript. Implicitly, the scenario
$\mathcal{S}_1$ as discussed above may be viewed a pump-probe scheme where the low frequency
components (pump) arrives earlier in time to first excite the LUMO, followed by
ionization mediated by the high frequency components (probe-pulse) arriving latter in time.
\begin{figure*}[tb]
  \centering
\includegraphics[width=0.95\linewidth]{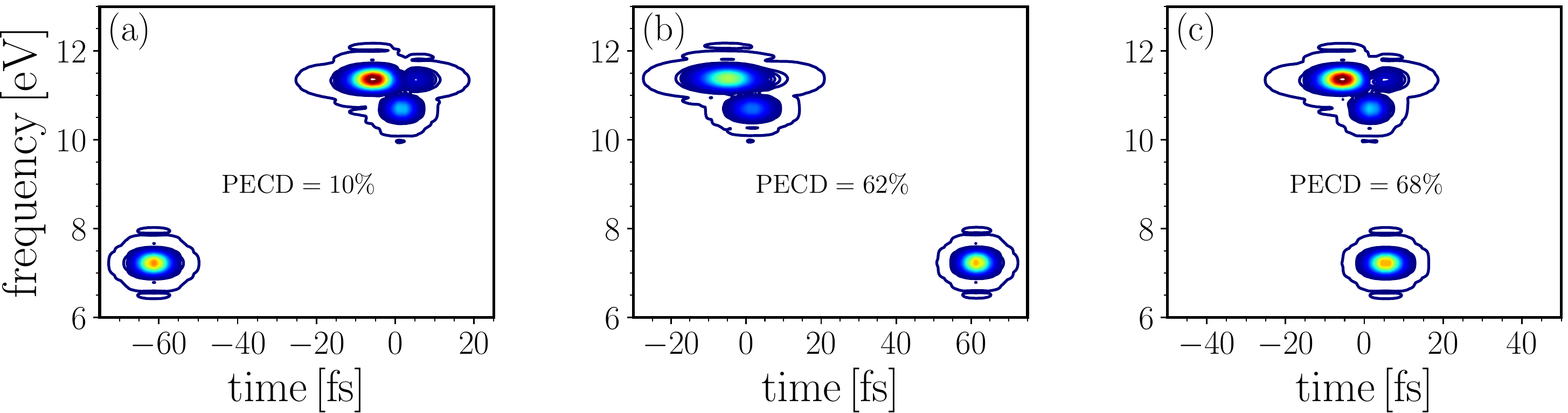}  
  \caption{Time-frequency distribution of the fully-optimized pulse for three
  different time delays between its low and high frequency components.  
  Pump-probe scheme corresponding to the scenario $\mathcal{S}_1$ is shown in (a). In this scenario, the
  pump and probe pulses, carrying the low and high frequency components,
  respectively,  are
  temporally separated. Only the two-photon pathway probing the LUMO
  participates in the photoionization process.
  Control scheme corresponding to the scenario $\mathcal{S}_2$ is shown in (b). Pump and probe pulses carrying
  high (pump)  and low (probe) frequency components are temporally separated.
  Maximum achievable PECD amounts to $62\%$ and the corresponding PECD contains
  the signature of all orbitals, except the LUMO. 
  Optimal scheme corresponding to the scenario $\mathcal{S}_3$, whereby pump and probe pulses carrying low and high frequency components
  overlaps in time is shown in (c). This scenario maximizes the PECD by exploting interference
  of all two-photon pathways including that through the LUMO.} 
\label{fig:WignerD}
  \end{figure*}

It is worth noticing, however, that exchanging  the temporal order of the pump and probe
pulses discussed above also lead to a resonantly enhanced two-photon ionization
scheme that allows the different contributing two-photon pathways to interfere exactly at
$6.5\,$eV. Such a scenario,  consisting in high (pump) and low (probe) frequency components arriving earlier and latter in time, respectively,
is referred to as $\mathcal{S}_2$,  and the corresponding two-photon ionization scheme 
is highlighted in blue in Fig.~2(f) in the manuscript. In detail, absorption of a photon with energy $\omega_{20}$ (high
frequency component)
resonantly populates the $\mathrm{LUMO}+1$. The photoelectron can \emph{then} be
promoted at $6.5\,$eV upon absorption of a photon with energy
$\omega_{10}+\delta\omega_1$ which is availated within the spectral width of
$\omega_1$ (low frequency component). Alternatively, according to the field
spectrum in Fig.~3(c), also the $\mathrm{LUMO}+3$
can be exited through the high frequency component $\omega_{30}$ contained in
the field,  and the
photoelectron can \emph{latter} be promoted at $6.5\,$eV upon absorption of $\omega_{10}-\delta\omega$, available within the
spectral width of $\omega_1$. Finally, same
two-photon mechanism can be retrieved by resonantly exciting the $\mathrm{LUMO+4}$.  

The main difference between the scenarios $\mathcal{S}_1$ and $\mathcal{S}_2$
lies in the different orbitals that are excited in the sequential
excitation-ionization scheme. In $\mathcal{S}_1$, only the $\mathrm{LUMO}$ is
exited by the pump pulsed and ionized afterwards. The PECD thus contain the signature of the $\mathrm{LUMO}$ alone. 

Conversely, in $\mathcal{S}_2$,  all
orbitals except the $\mathrm{LUMO}$ are excited by the pump pulse containing
the high frequency components.  Thus, after ionization, 
the PECD contains the signature of all orbitals, except that of the
$\mathrm{LUMO}$.

Finally, the third possible scenario, $\mathcal{S}_3$, consists in both low and high frequency
components of the optimized spectrum sharing a given time window, this is, pump
and probe pulses overlapping in time. The main interest in overlapping pump and
probe pulses
lies in the fact that,  in contrast to $\mathcal{S}_2$, also the $\mathrm{LUMO}$ can be
excited  by the pulse  that now contains not only high-frequency
components to excite the  $\mathrm{LUMO}+j$ orbitals, for $j\ge 1$, but also the required low frequency 
components to excite the $\mathrm{LUMO}$. Consequently, the signature of
the $\mathrm{LUMO}$ along with higher-lying orbitals is imprinted in the PECD. It remains to be seen, however,  which
scenario, namely $\mathcal{S}_1$, $\mathcal{S}_2$ or $\mathcal{S}_3$ indeed corresponds to  
the optimized two-photon scheme leading to a PECD of $68\%$ at $6.5\,$eV.

In order to disentangle the optimal scenario from all three possible cases
discussed above, we have introduced a time delay, $\tau$,  between the low and high
frequency components, which will be varied from negative to positive values.
Positive (negative) delays correspond to the high-frequency components arriving before (after) the low-frequency components. 
Analyzing the PECD as a function of $\tau$ as presented in Fig.~3(c)
in the manuscript allow us to retrieve the optimal scenario. In fact, a time-frequency analysis
of the optimized pulse for the optimal time-delay yielding to $68\%$, tells us
about whether the low and high frequency components overlap in time ($\mathcal{S}_3$)
or whether they are temporally separated, i.e. sequential pump-probe scheme ($\mathcal{S}_2$ or $\mathcal{S}_3$)\footnote{Although 
there is no need to introduce an external delay $\tau$ to evaluate the Wigner
distribution function of the fully-optimize pulse, i.e, the optimized pulse already contains the optimal
time-delay, this procedure allows us to confirm  the role of the pump and probe
pulses in terms of the orbitals being excited prior to the ionization step.}. 
Also note that since both low and frequency components are spectrally separated (not overlapping), the power spectrum remains unaltered.

In Fig.~3(c) in the manuscript, we show the PECD as a function of the time-delay
$\tau$. The section highlighted in red in Fig.~3(c) corresponds to the first
scenario $\mathcal{S}_1$. Figure~\ref{fig:WignerD}(a) shows the
Wigner distribution function for a time-delay corresponding to 
$\mathcal{S}_1$. As discussed before, only the $\mathrm{LUMO}$ participate in the sequential two-photon ionization process, 
as the low frequency components arrive earlier in time. This is also confirmed by 
Fig.~3(c) (highlighted in red) in the manuscript. In fact, the corresponding PECD  exhibits 
no oscillation as a function of the time-delay, which
is explained by the absence of interfering  two-photon ionization pathways at $6.5\,$eV, as only the LUMO contributes
to the PECD. The steady value for the PECD amounts to $10\%$.

For positive time delays (highlighted in blue in Fig.~3(c) in the main text) the
pump and probe pulses are now switched. This is, the high frequency components
arrive earlier in time, as verified by the Wigner distribution function shown
in Fig.~\ref{fig:WignerD}(b), which shows the time-frequency 
distribution for a time delay corresponding to $\mathcal{S}_2$. In this scenario, the pump 
pulse --now containing the high-frequency components of the spectrum--  creates a wavepacket defined by
a superposition of all orbitals except the $\mathrm{LUMO}$, which evolves under
the field-free Hamiltonian. After a time
delay $\tau$, the probe pulse containing the low frequency components  ionizes the prepared wavepacket that has gained a
given time delay-dependent phase. The signature of
the interference among the contributing two-photon pathways can be seen in 
Fig. 3(c) (highlighted in blue) in the manuscript. The corresponding PECD exhibits
an oscillatory behavior as a function of the time-delay, associated with the
time delay-dependent phase accumulated by the prepared wavepacket during the
time-delay between the pump and probe pulses.
According to Fig.~3(c) in the manuscript, the maximum
PECD for this sequential pump-probe scheme accounts for $55\%$\footnote{further
optimization within this pump-probe scheme leads to a PECD of $62\%$, as
discussed in the manuscript.}, which is not 68\% found for the fully optimized
pulse. This means the optimal scheme does not correspond to a pump-probe scheme
whereby low and high and frequency components are temporally separated.

Only for a specific time delay of $8.2\,$fs, the optimal PECD (68\%) is retrieved,
cf.~Fig.~3(c) (highlighted in yellow) in the manuscript. The corresponding time-frequency distribution 
is depicted in Fig.~\ref{fig:WignerD}(c).  It shows that 
the optimal case corresponds to both ``pump'' and ``probe'' pulses 
sharing a particular temporal window. In other
words, the optimal scenario correspond to low and high frequency components
overlapping in time, i.e. $\mathcal{S}_3$. The stronger PECD of 68\% compared to that obtained in
the framework of $\mathcal{S}_2$ (55\%) can be explained 
by exploiting interference of all the pathways, including 
the two-photon ionization through the LUMO\footnote{since the low frequency is now also
available at the same time that the higher frequency components}, which now also contribute to the
PECD. The respective ionization scheme is depicted in Fig.~2(f) (highlighted in yellow) in the
manuscript.  

As alluded to above, we conclude that the highly efficient control of PECD is achieved 
via various (1+1$^\prime$) REMPI pathways
that probes different intermediate states to constructively interfere at
a common photoelectron energy. By introducing a time delay between the low and
high frequency components of the optimized field, we have shown  that 
the control mechanism based on multiple even-parity interference pathways 
presented in this work outperforms the standard sequential pump-probe scheme as it 
maximizes the number of molecular states that 
constructively contribute to the dichroism at an optimal photoelectron kinetic energy.

\section{\textbf{\sffamily{Maximizing the PECD at a specific photoelectron energy using the multiple $(1^\prime+1)$-REMPI scheme}}}
 \label{sec:ResultsAt10eV}

\begin{figure*}[tb]
  \centering
\includegraphics[width=0.90\linewidth]{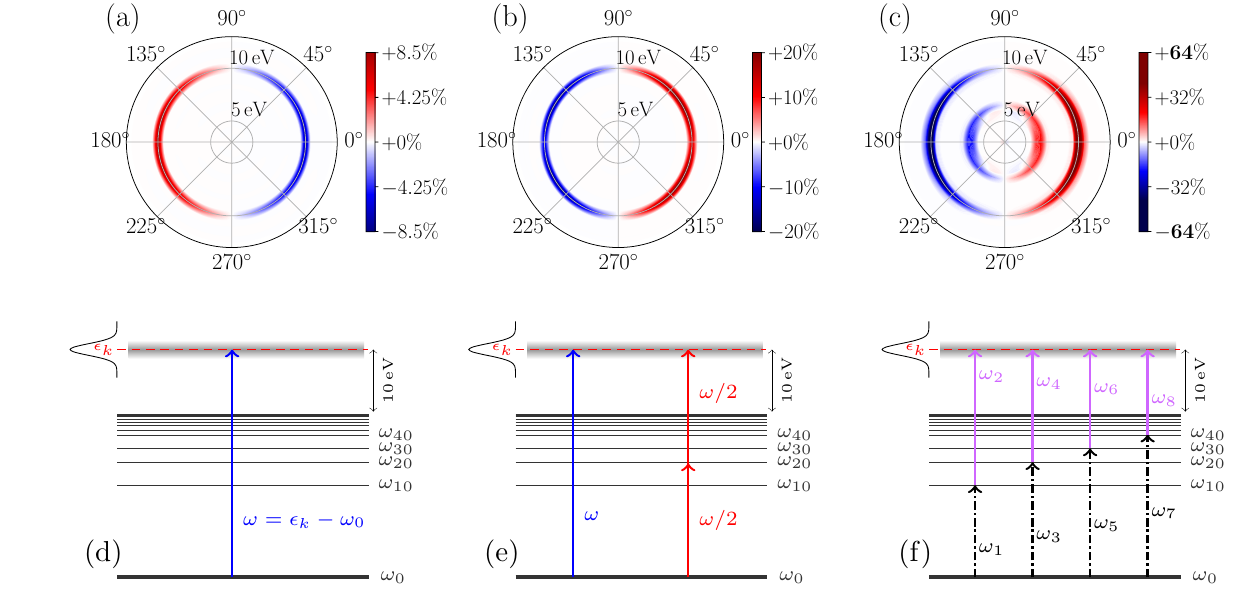}  
  \caption{PECD obtained with a reference field driving one-photon ionization only (a) , optimized bichromatic
  $(\omega,2\omega)$ pulse  (b), and 
making use of interference in even-parity two-photon pathways (c). For all three
schemes, the corresponding maximum PECD is obtained at the same final photoelectron
  kinetic energy $\epsilon^*_k$ of $10$ eV.
Ionization and control schemes for the guess and optimized      
$(\omega,2\omega)$ pulses are shown in panels (d) and (e), respectively.
The control mechanism for even-parity pathway interference (f)
is based on   
probing different intermediate states that interfere constructively at a common
continuum photoelectron energy within the spectral bandwidth.  
}
\label{fig:Figure1.suppmat}
  \end{figure*}
As a complement to the PECD study with a peak 
of the photoelectron distribution at an 
energy of $6.5\,$eV discussed in the main text, we provide here the optimization
results for a peak at $10\,$eV. The latter corresponds to the 
photoelectron energy for which the largest PECD, of about 20\%, 
is obtained when using a bichromatic $(\omega,2\omega)$ pulse.
Starting from a reference field driving one-photon ionization only, Fig.~\ref{fig:Figure1.suppmat}(a) displays the
 corresponding single-photon PECD, with a maximum of   
$8.5\%$ at a photoelectron kinetic energy
 $\epsilon^*_k$ of
$10$ eV. The one-photon ionization scheme                                      
is
schematized in Fig.~\ref{fig:Figure1.suppmat}(d). Note that the PECD value can also be
retrieved from Figure~1(b) in the main text (``one-photon'' regime) at a photon energy of  $\omega=21.88$ eV (solid-blue lines).

The PECD resulting from the optimized bichromatic $(\omega,2\omega)$ field  
is shown in Fig.~\ref{fig:Figure1.suppmat}(b). 
In this case, the second harmonic  and fundamental photon energies correspond to  $\omega=21.88$ eV and $\omega/2=10.94$ eV, respectively. 
The orbital energies 
for the $\mathrm{LUMO}$ and $\mathrm{LUMO+1}$  are equal to $-4.80$ and $-0.97$ eV, respectively.
In contrast to the case of a photoelectron energy $\epsilon^*_k=6.5$ eV discussed in the main text, where the excitation 
due to  absorption of a photon with
energy corresponding to the fundamental was found  to exactly lie in between the orbital 
energies of  the $\mathrm{LUMO}$ and $\mathrm{LUMO+1}$ (at $-2.69$ eV),
cf. main text, here, $\omega_0+\omega/2=-0.93$ eV, to be compared to  the energy of the
$\mathrm{LUMO+1}$, i.e., $\omega_0+\omega_{20} =-0.97$ eV. Consequently, for an optimal
photoelectron energy $\epsilon^*_k=10$ eV, the efficiency of the two-photon ionization 
is enhanced compared to that at $6.5$ eV discussed in the main text. As a result, 
a larger magnitude for the PECD is found ($20\%$ instead of $14\%$). The corresponding ionization
scheme is depicted in Fig.~\ref{fig:Figure1.suppmat}(e).

Finally, the PECD resulting from the multiple two-photoionization scheme converging at $10$ eV is shown in
Fig.~\ref{fig:Figure1.suppmat}(c). A maximum $\mathrm{PECD}=64\%$ exactly at $10$ eV is obtained.    
Analogously to the enhancement mechanism at $6.5$ eV discussed in the main text, the
physical mechanism responsible for a PECD of $64\%$ at $10$ eV  is also based on multiple
$(1^\prime+1)$-REMPI process probing different molecular orbitals
and coherently converging at the same final photoelectron
energy. Only two-photon processes participate\footnote{main contribution arising from the first four
LUMO orbitals. The transition dipole moments from the HOMO to these orbitals
depends on the molecular orientation with respect to the electric field
polarization direction. The square of the isotropically averaged transition dipole moments (atomic units) correspond to $0.28$, $0.16$, $0.20$ and
$0.02$, respectively.}, with  no contribution from one-photon ionization
and interference between opposite-parity ionization pathways  at $10$ eV. 
The corresponding ionization scheme is as depicted in Fig.~\ref{fig:Figure1.suppmat}(f). 

For the optimization based on the multiple ($1^\prime+1$)-REMPI (coherently) converging at $10$
eV,  the photon energies $\omega_j=\omega_{j0}$ together with $\omega_{j+n} = \epsilon^*_k - \omega_{j,0} + |\omega_0|$, for
$j=1,n$ were kept fixed, whereas the remaining control variables,
cf.~Eq.~(9) in the manuscript, were optimized. Here,  $n$
refers to the number of two-photon pathways, each one probing $n$ different orbitals, to be accounted for.
Figure~\ref{fig:Figure1.suppmat}(f) thus schematizes the (multiple) two-photon ionization
scheme for $n=4$, i.e.  probing $n=4$ different intermediate states but converging at the
same photoelectron energy, within the spectral width.

\begin{figure*}[tb]
\centering
\includegraphics[width=0.90\linewidth]{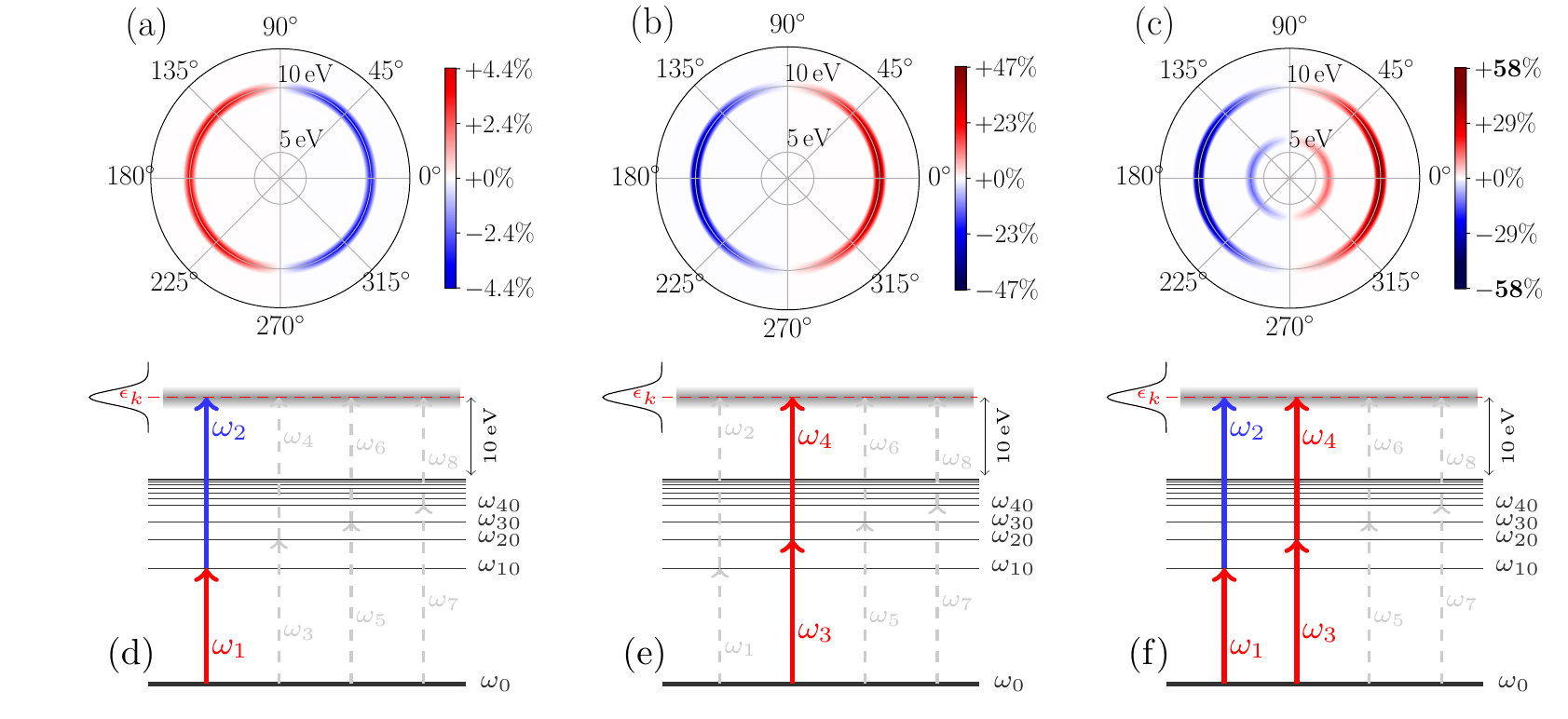}  
  \caption{Optimized anisotropic response obtained within the multiple $(1^\prime+1)$-REMPI
  scheme using one, panels (a) and (b) and two (c)  distinct 
  two-photoionization pathways. In panel (a), the REMPI process consist in
  resonant excitation of the $\mathrm{LUMO}$ orbital, followed by ionization out of the
  resonantly excited state, as schematized in panel (d). The resulting optimized PECD 
  corresponds to  $4.4\%$. Excitation of the ${\mathrm{LUMO+1}}$
  instead of the $\mathrm{LUMO}$ (e)
  in the REMPI process results in a significant larger PECD effect $(47\%)$, at the same
  photoelectron kinetic energy ($10$ eV).  Combination of both schemes, i.e., of the two photoionization pathways, 
  enhances   the PECD to $(58\%)$ (c) . In all three cases, the
  optimization were carried independently. Incorporation of additional pathways increases the PECD effect even more, e.g.
  $\mathrm{PECD}=64\%$ for four two-photoionization pathways converging at the
  same photoelectron energy as displayed in Fig.~\ref{fig:Figure1.suppmat}(c) and Fig.~\ref{fig:Figure1.suppmat}(f).} 
\label{fig:Figure2.suppmat}
\end{figure*}
Reducing the number $n$ of contributing two-photon pathways  results in a decrease of
PECD (abs. value) at its maximum peak distribution. 
Figure~\ref{fig:Figure2.suppmat} shows the optimized PECD                         
at $10$ eV obtained with $n=1$ (panels a-d)
and  $n=2$ (panels c, f)                                       
distinct two-photon
ionization pathways. For Fig.~\ref{fig:Figure2.suppmat}(a), the pulse is
parametrized according to Eq.~(9) in the main text with only two frequencies
components: the frequencies 
$\omega_1$ and $\omega_2$, as indicated in Fig.~\ref{fig:Figure2.suppmat}(d),
are kept fixed and the remaining pulse parameters are optimized. Analogously, for
Fig.~\ref{fig:Figure2.suppmat}(b), the optimized field consists in a pulse with
two (fixed) frequencies components $\omega_3$ and $\omega_4$, as indicated in
Fig.~\ref{fig:Figure2.suppmat}(e), and the remaining
pulse parameters are treated as optimization variables. Finally, for
Fig.~\ref{fig:Figure2.suppmat}(c), i.e. $n=2$, the optimized pulse consist of four (fixed)
frequencies, namely $\omega_1, \omega_2,\omega_3, \omega_4 $, as indicated in
Fig.~\ref{fig:Figure2.suppmat}(f) . In all
three scenarios the optimizations were performed independently. 

The maximum chiral response is obtained at the same final photoelectron energy for all three cases. However, from
Fig.~\ref{fig:Figure2.suppmat}(a) and (b), it is apparent that both two-photon
ionization pathways  imprint their own signature to the PECD, as the
resulting magnitude of PECD changes dramatically depending on the two-photon pathway.  Consequently, 
the contributing two-photon pathway plays a critical
role for maximizing the PECD. Specifically, for a two-photon pathway probing only the
LUMO orbital followed by ionzation, cf.~Fig.~\ref{fig:Figure2.suppmat}(d),  the PECD amounts to $4.4\%$ only.  In contrast,  the ionization pathway 
probing the $\mathrm{LUMO+1}$ with subsequent ionization, cf.~Fig.~\ref{fig:Figure2.suppmat}(e), results in
an orientation averaged PECD of $47\%$. When the optimization is performed using the combined two-photon pathways, cf.~Fig.~\ref{fig:Figure2.suppmat}(f), the
PECD is enhanced, in a synergetic manner,  to $58\%$, as shown in Fig.~\ref{fig:Figure2.suppmat}(c),       
indicating the coherent nature of the interference process. Furthermore, adding two more pathways, 
such that also the $\mathrm{LUMO+2}$ and $\mathrm{LUMO+3}$ are probed, as indicated in Fig.~\ref{fig:Figure1.suppmat}(f), 
results in the optimal scheme for a photoelectron energy $\epsilon^*_k$ of $10$ eV,  with $\mathrm{PECD}$ equal to $64\%$, as shown in Fig.~\ref{fig:Figure1.suppmat}(c). The
optimal multiple $(1^\prime+1)$ REMPI scheme, for $10$ eV, therefore consists in
four different two-photon ionization pathways, i.e. $n=4$, converging
and constructively interfering at the same photoelectron energy, as depicted in Fig.~\ref{fig:Figure1.suppmat}(f).

\begin{figure*}[tb]
\centering
  \includegraphics[width=0.99\linewidth]{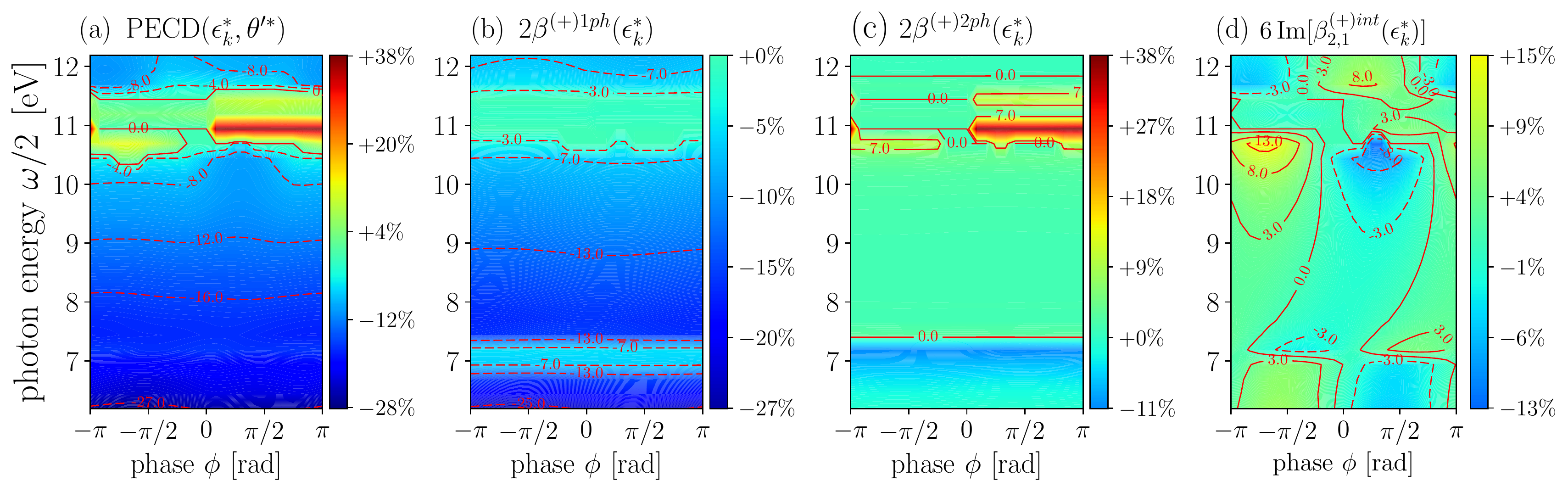}  
  \caption{PECD (a) as a function of the photon energy (fundamental) and
  relative phase between the fundamental and second harmonic obtained with a bichromatic ($\omega,2\omega$) pulse. Maximum Contribution (at their respective $\theta^{*}_{k^\prime}$) 
  of the one- and two-photon ionization pathways are shown in panels (b) and
  (c), respectively. Panel (d) depicts the maximum contribution of the interference
  between one- and two-photoionzation pathways at $\theta^*_{k^\prime}=\pi/2$ (maximum
  contribution for the interference term). The  large magnitude of PECD ($38\%$) is
  solely due to two-photon ionization pathways (c) whereas  in the lower part of panel
  (a), the large magnitude of PECD ($-27\%$) is mainly due to one photon process (b).
  The interference term does not contribute to these values as both  
  are obtained at $\theta^{*}_{k^\prime}=0$ (modulo $\pi$),  and the contribution from the interference
  between one- and two-photon ionization pathways vanish at these angles, cf.~Eq.\eqref{eq:FinalPecdEquation}. The main
  contribution from the interference term is obtained for the fundamental $\omega/2=10.94$ eV 
  and second harmonic $\omega=21.88$ eV, cf. panels (a) and (d). The resulting 
   (total) $\mathrm{PECD}$ of $20\%$ is found at a photoelectron kinetic   
  $\epsilon^*_k$ of $10$ eV, with an angle $\theta^{*}_{k^\prime}$ of $\pi/2$ rads, cf.~Fig.~\ref{fig:Figure1.suppmat}(b)}
\label{fig:Figure3.suppmat}
\end{figure*}
Finally, Fig.~\ref{fig:Figure3.suppmat} presents a photon energy vs. relative
phase correlation map for the PECD obtained with a bichromatic $(\omega, 2\omega)$ pulse
with fixed FHWM and intensities, corresponding to that of the optimized bichromatic ($\omega,2\omega$) field
sketched in Fig.~\ref{fig:Figure1.suppmat}(e).
The total PECD and anisotropy parameters defined in Eq.~\eqref{eq:FinalPecdEquation} 
for the first and second order process --and their interference-- are shown in Fig.~\ref{fig:Figure3.suppmat}, panels (a)-(d), respectively, as a function of the
photon energy (fundamental) and relative phase between the fundamental and second
harmonic. The PECD in Fig.~\ref{fig:Figure3.suppmat}(a) corresponds to the
maximum PECD (abs. value) extracted from the PECD distribution in Eq.~\eqref{eq:FinalPecdEquation},  obtained  at the particular  
photoelectron kinetic energy $\epsilon^*_k$ and direction
$\theta^{*}_{k^\prime}$, at which the PECD is maximal, i.e.,  
$|\mathrm{PECD}(\epsilon^*_k, \theta^*_{k^\prime},\pi/2)|\ge |\mathrm{PECD}(\epsilon_k, \theta_{k^\prime}, \pi/2)|$, 
$\forall\, (\epsilon_k,\theta_{k^\prime})$.

It is worth noting, however, that interpretation of Fig.~\ref{fig:Figure3.suppmat}(a) must be taken with precaution
since a large magnitude of PECD obtained with a bichromatic $(\omega,2\omega)$ field does not necessarily mean that interference
between one-photon and two-photon pathways plays a significant role.
In fact, from Figs.~\ref{fig:Figure3.suppmat}(a) and (c), 
it is apparent that the large magnitude of positive PECD, of $\approx 38\%$, is almost exclusively
due to two-photon ionization pathways. This finding  is further confirmed by removing the
second harmonic from the bichromatic field. Note also that this large value
can already be retrieved from the red curve in Figure~1(b) of the main text, corresponding to the two-photon regime.

Conversely,  in the lower part of
Fig.~\ref{fig:Figure3.suppmat}(a), the large  negative PECD, of $\approx -27\%$, is mainly  due to a one-photon process, which can be
seen by comparing panels (a) and (b) in Fig~\ref{fig:Figure3.suppmat}. Very importantly, 
the angle at which the PECD is maximal, $\theta^{*}_{k^\prime}$, must, of course,  also be taken
into account. In fact, 
although the interference term, $\beta^{int}_{2,1}(\epsilon^{*}_k)$,  shown in Fig~\ref{fig:Figure3.suppmat}(d) does
not vanish for the pulse parameters yielding a $\mathrm{PECD}$ equal to $-27\%$, a further inspection of the direction $\theta^*_{k^\prime}$ of 
photoelectron emission at which the $\mathrm{PECD}$ is $-27\%$ reveals that  $\theta^{*}_{k^\prime}$ corresponds to $0$ (and $\pi$).                          
For $\theta_{k^\prime}$ equal to $0$ (nor $\pi$),  the 
interference term between opposite-parity photoionzation pathways cannot contribute at all, cf. Eq~\eqref{eq:FinalPecdEquation}.
The same applies for the maximum
PECD of $+38\%$, previously discussed. 

Instead, the maximum contribution from the interference term, or more rigorously speaking, the
best compromise between all terms in Eq.~\eqref{eq:FinalPecdEquation} to
maximize the PECD while requiring       
the interference term to play a non-negligible role                              
is found for $\omega/2=10.9$ eV (fundamental) for 
a total $\mathrm{PECD}$ of $20\%$ at a photoelectron kinetic energy distribution of 
$10$ eV., cf.~Fig.~\ref{fig:Figure3.suppmat} (a) and (d), with an angle $\theta^{*}_{k^\prime}$ equal to $\pi/4$       
(modulo $\pi$), cf.~Fig.~\ref{fig:Figure1.suppmat}(b).

\section{\textbf{\sffamily{Input file parameters for }} \texttt{MOLPRO} }
\label{sec:section4}
\texttt{\hspace{-0.3cm}$\#$ MOLPRO Program package~\cite{werner2012molpro,werner2012molprowires} input file}

\texttt{***,CHFClBr gs}\\
\texttt{RCH = 1.8 ANG}\\
\texttt{TANG = 109.471220634491 DEGREES}\\
\texttt{RCCl = 2.4 ANG}\\
\texttt{RCBr = 2.5 ANG}\\
\texttt{RCF = 2.5 ANG}\\
\texttt{cartesian}
\texttt{geometry=\{}\\
  \texttt{C1;}
  \texttt{H1,C1,RCH;}
  \texttt{Cl2,C1,RCCl,H1,TANG;}\\
  \texttt{Br3,C1,RCBr,H1,TANG,Cl2,TANG,1;}\\
  \texttt{F4,C1,RCF,H1,TANG,Cl2,TANG,-1;}\\
\texttt{\}}\\
\texttt{basis=avdz}\\
\{ \texttt{rhf;}\\
\texttt{wf,68,1,0;\}}\\
\texttt{put, molden, chfclbr.molden;}\\
\texttt{optg;}

\section{\textbf{\sffamily{Input file parameters for}}\,\,\,\texttt{\lowercase{e}P\lowercase{oly}S\lowercase{cat}}}
\label{sec:section5}
\texttt{\hspace{-0.3cm}$\#$ ePolyScat Program package~\cite{GianturcoJCP94,NatalenseJCP99} input file}\\
\texttt{\hspace{-0.3cm}$\#$ input file CHFClBr photoionization}\\
\texttt{$\#$\, and matrix elements calculation}\\
\medskip

\texttt{\hspace{-0.35cm}LMax  120}\\         
\texttt{EMax  50.0}\\      
\texttt{EngForm}\\          
\texttt{0 0}\\             
\texttt{FegeEng 10.98}\\    
\texttt{LMaxK 60}\\ 
\texttt{NoSym}\\
\texttt{InitSym 'A'}\\                               
\texttt{InitSpinDeg 1}\\                                    
\texttt{OrbOccInit 2 2 2 2 2 2 2 2 2 2 2 2 2 2 2 2 2 2 2 2 2 2 2 2 2 2 2 2 2 2 2 2 2 2}\\ 
\texttt{OrbOcc  2 2 2 2 2 2 2 2 2 2 2 2 2 2 2 2 2 2 2 2 2 2 2 2 2 2 2 2 2 2 2 2 2 1}\\ 
\texttt{SpinDeg 1}\\                                                                     
\texttt{TargSym 'A'}\\                                                                   
\texttt{ScatContSym 'A'}\\                                                               
\texttt{ScatSym 'A'}\\                                                             
\texttt{TargSpinDeg 2}\\                                                                 
\texttt{Convert 'chfclbr.molden' 'molden2012'}\\
\texttt{FileName 'MatrixElements' 'MatEleOrbs.idy' 'REWIND'}\\
\texttt{GetBlms}\\
\texttt{ExpOrb}\\ 
\texttt{GenFormPhIon}\\ 
\texttt{GetPot}\\
\texttt{DipoleOp}\\
\texttt{PhIonN 0.1 0.1 160}\\

\end{document}